\documentclass[a4paper,11pt]{article}
\usepackage{amsfonts}

\usepackage{graphicx}
\graphicspath{{Figures/}}
\usepackage{multicol}
\usepackage{amsmath}
\usepackage{amssymb}
\usepackage{latexsym}
\usepackage{color}
\usepackage{float}
\usepackage{cite}
\usepackage{makeidx}
\usepackage{colortbl}
\usepackage{csquotes}
\usepackage[squaren]{SIunits}
\usepackage{braket}
\usepackage[colorlinks,linkcolor=red,citecolor=green]{hyperref}
\usepackage[textfont={footnotesize,sf},labelfont={color=blue,bf,sf},labelsep=endash]{caption}
\usepackage[position=top,labelfont={color=blue,bf,sf}]{subfig}
\usepackage{bm}

\usepackage[title]{appendix}

\begin{document}
	\begin{titlepage}
		\setcounter{page}{1}
		\renewcommand{\thefootnote}{\fnsymbol{footnote}}

		\begin{center}
			
			{\Large \bf {{Strain Effect on Transmission in Graphene Laser Barrier}}}

			\vspace{6mm}
			
	{\bf Hasna Chnafa\footnote{\sf {\color{magenta} chnafa.h@ucd.ac.ma}}}$^{a}$, {\bf Miloud Mekkaoui\footnote{\sf {\color{magenta} miloud.mekkaoui@gmail.com}}}$^{a}$, {\bf Ahmed Jellal\footnote{\sf {\color{magenta}a.jellal@ucd.ac.ma}}}$^{a,b}$ and {\bf Abdelhadi Bahaoui}$^{a}$

			\vspace{5mm}
			
			{$^{a}$\em Laboratory of Theoretical Physics,  
				Faculty of Sciences, Choua\"ib Doukkali University},\\
			{\em PO Box 20, 24000 El Jadida, Morocco}

			{$^{b}$\em Canadian Quantum  Research Center,
				204-3002 32 Ave Vernon, \\ BC V1T 2L7,  Canada}

		\vspace{3cm}	
\begin{abstract}
We investigate the strain effect along armchair and zigzag directions on the tunneling transport of Dirac fermions in  graphene laser barrier through a time dependent  potential along $y$-axis. Our system is composed of three regions and the central one is subjected to a deformation of strength $S$. Based on  Dirac equation and  the Floquet approach, we determine 
the eigenvalues and  eigenspinors for each region. Using  the boundary conditions at interfaces together with the transfer matrix method we identify the transmission {in the different Floquet sideband states as function} of the physical parameters. 
{In the strainless case, we show that the transmisson of  central band decreases for smaller values of the barrier width and   rapidly oscillates with different amplitude for  larger ones. Whereas the transmission for the first sidebands increases from zero and shows a damped oscillatory profile. It is found that {the number of oscillations} in all transmission channels {reduces 
{with} increasing the strength of armchair strain but becomes more important by switching the deformation to zigzag}. Moreover, it is observed the appearance of Fano type resonance peaks by altering the amplitude and the frequency of the laser field}.			
\end{abstract}	
\end{center}

\vspace{5cm}

\end{titlepage}
		\section{Introduction}
 Graphene is an intriguing subject of modern physics \cite{q1} and remains  among the most important discovery  
in material science \cite{q1,q2}. Its band structures are described by a low-energy theory similar to the massless Dirac-Weyl fermions. Graphene exhibits a number of unusual electronic and mechanical properties \cite{q5} such as, Hall effect \cite{q6,q7,q8}, elastic strain engineering \cite{q9,q10,q11,q12}, Klein tunneling \cite{q13,q6,q15} and so on. It is showed that the tunneling effect has already observed experimentally \cite{qa15} in graphene system and has been investigated and achieved many important results in the single \cite{q16} or double potential barriers \cite{q17}. 
On the other hand, when a mechanical constraint is applied to graphene, it creates a deformation leading to a change of its properties 
{and thus improve its technology functionality. For example, a band gap opening was obtained using a uniaxial constraint \cite{q9,S2}. Moreover, among the implications induced by the most interesting constraints, one can cite the experimental observation of a spectrum resembling the Landau levels in graphene under strain \cite{S3,S4}, which was created earlier employing gauge fields \cite{S5}}. To this end, it is showed that the tensional strain can be  used to control the electronic and optical properties of graphene. Also strain can modify the Dirac points, which causes Dirac fermions having asymmetric effective Fermi velocities in either monolayer \cite{ref25,a5} or bilayer graphene \cite{a7}.

{\color{black} On other hand, the advances in laser physics and microwave technology carried out over the past decades become an important area of research. They allowed to get modified physical properties of driven system, which resulted in many fundamental effects in various nanostructures, including the quantum well in conventional semiconductor \cite{Xu,Xv,gt}, {the quantum rings \cite{ly1,ly2}}, the monolayer and bilayer graphene \cite{Xu1,Xu2,strain,Xu3,XU1,XU2} as well as others}. {The study of the external laser field assisted electronic transport through graphene in the framework of the Floquet theory is an interdisciplinary branch of physics. Indeed, it  is very crucial to understand the physics behind the light-matter coupling, which has  been investigated by different groups. For this, we cite transmission of electron through monolayer graphene laser barrier \cite{f1}, photon induced tunneling of electron through a graphene electrostatic barrier \cite{Xu4}, beating oscillation and Fano resonance in the laser assisted electron transmission through graphene $\delta$-function magnetic barriers \cite{Xu5}.} {\color{black}Although there are  some works carried out on laser assisted in graphene have been declared, the presence of strain effect under a laser field through an electrostatic barrier is an attractive problem to be studied, which constitutes the subject of our paper}.
		


Motivated by the results obtained in \cite{f1}, we theoretically investigate the strain effect on the tunneling transport of Dirac fermions 
in graphene subjected to a linearly polarized laser field. In the first stage, we solve the Dirac equation to analytically obtain the eigenvalues and corresponding eigenspinors for each regions. Using the boundary condition with transfer matrix approach and the Floquet theory, we determine the transmission 
in terms of the physical parameters. 
 Subsequently, we numerically study the  effect of strain along armchair and zigzag directions on the transmission 
for the sideband states including the central one. 
 %
 Consequently, we show that in the case of laser irradiation the effect of strain causes some amendments on the transmission behavior for armchair deformation while it produces a admirable impact for the zigzag one. We conclude that the strain amplitude can be used as a key component to control tunneling properties of our system. 

The outline of the present paper is as follows. In {\color{red}\textbf{Sec. 2}}, we set the mathematics formalism based on the Hamiltonian describing
our system
and build the solutions of energy spectrum using the Floquet theory.   Matching the eigenspinors 
at interfaces and apply the transfer matrix approach, we explicitly determine  the transmission probability for all energy modes in {\color{red}\textbf{Sec. 3}}.  We numerically analyze and discuss our results under suitable choices of the physical parameters in {\color{red}\textbf{Sec. 4}}.  Finally, we conclude our results.	

\section{Theoretical formalism}
\begin{figure}[H] \qquad\qquad\quad	
\centering
\includegraphics[scale=0.46]{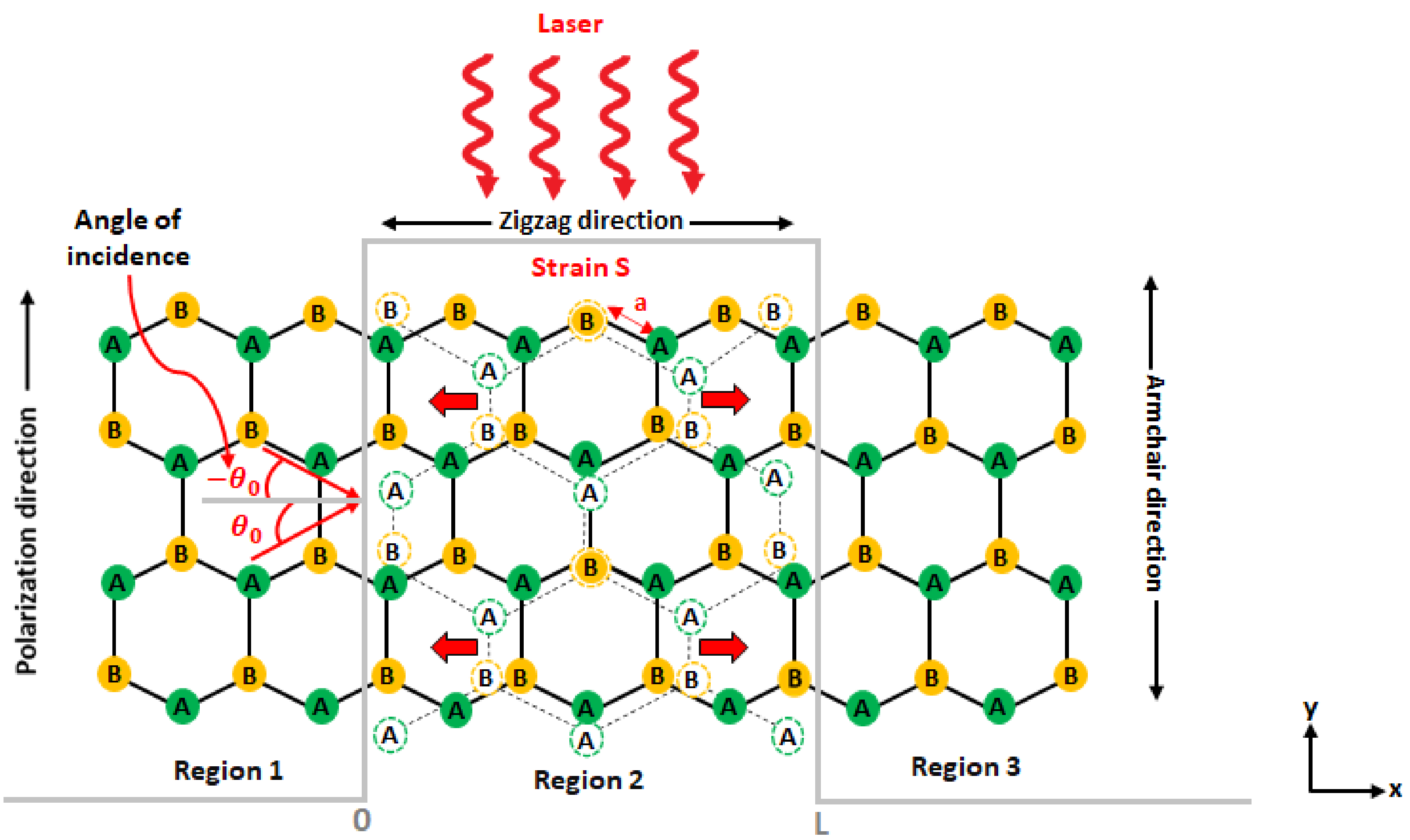}
\caption{\sf{(color online) {
{Schematic diagram of the normal/strained/normal}		graphene irradiated by linearly polarized field for tensile strain along zigzag ($x$-axis) and armchair ($y$-axis) directions {with the solid and dashed circles denote sublattices  \text{A} (green) and \text{B} (yellow) in undeformed and deformed configurations, respectively}}}.}\label{db.5}
\end{figure}

{We consider a system made of  graphene subjected to 
a}	
laser field of frequency $\omega$, which is linearly polarized along the $y$-axis. Our system contains three regions labeled by $j=1,2,3$, such that the \textbf{region 1} and \textbf{3} are assumed to be the normal graphene stripes, whereas the \textbf{region 2}  is the strained stripe with the strain exerted along armchair and zigzag directions as showed in {\color{blue}\textbf{Figure}} $\ref{db.5}$.  Before to proceed further, let us notice that in the presence of a laser field,
the incident electrons of angle $\theta_{0}$ and energy $\mathcal{\varepsilon}$ are reflected with energies $\mathcal{\varepsilon}+n\hbar\omega$ and forming the angles $\pi-\theta_{n}$ after reflection and $\theta_{n}$ after transmission, where $n=0,\pm 1,\pm 2, \cdots$ represents the modes which generated by the oscillations.

Considering the electron-field interaction  the electronic properties of dressed electrons in our system can be described by the Hamiltonian
\begin{equation}\label{H1}
	\mathcal{H}=\begin{pmatrix}
			0&v_{x} \left(p_{x}-eA_{x}\right)-iv_{y} \left(p_{y}-eA_{y}\right)  \\
	 v_{x} \left(p_{x}-eA_{x}\right)+iv_{y} \left(p_{y}-eA_{y}\right) & 0
	\end{pmatrix}
\end{equation}
where $\textbf{p}=(p_x,p_y)$ is the momentum operator
and the vector potential $\textbf{A}=\left(A_{x},A_{y}\right)$ associated to the dressing field. The effective velocities
$v_x$ and $v_y$ 
are   given by the tensional strain \cite{ref25,a6,a5,a7,a8} defined within the strip $0\leq x\leq L$ but $v_x(S=0\%)$, $v_y(S=0\%)\rightarrow v_{F}$ elsewhere. The strain mentioned is assumed to be applied along armchair or zigzag directions with the corresponding effective velocities are written as
\begin{eqnarray}\label{xt1}
	v_{x}=\frac{\sqrt{3}}{2\hbar}a(1-\sigma
	S)\sqrt{4t_{1}^{'2}-t_{3}^{'2}}, \qquad  
	v_{y}=\frac{3}{2\hbar}a(1+ S)t_{3}^{'}
\end{eqnarray} when strain exerted along armchair direction and
\begin{eqnarray}\label{xt2}
	v_{x}=\frac{\sqrt{3}}{2\hbar}a\left(1+S\right)\sqrt{4t_{1}^{'2}-t_{3}^{'2}},\qquad
	v_{y}=\frac{3}{2\hbar}a(1-\sigma S)t_{3}^{'} 
\end{eqnarray} for zigzag direction. According to the elasticity theory \cite{yu}, the application of the tensional strain deforms the three nearest neighbor vectors for different strain magnitudes such as
\begin{eqnarray}\label{xtzr}
&&|d^{'}_{1}|_{(A)}=|{d}^{'}_{2}|_{(A)}=a\left( 1-\frac{3}{4}\sigma S+\frac{1}{4}S\right), \qquad |d^{'}_{3}|_{(A)}=a\left( 1+S\right)\\\label{xtt}
&&|{d}^{'}_{1}|_{(Z)}=|d^{'}_{2}|_{(Z)}=a\left(
1+\frac{3}{4} S-\frac{1}{4}\sigma S\right), \qquad
|d^{'}_{3}|_{(Z)}=a\left( 1-\sigma S\right)
\end{eqnarray} where $\sigma\approx 0.165$ is the Poisson ratio for graphene, $S$ is the strain magnitude, $a=0.142$ \text{nm} is the distance between neighboring atoms in pristine graphene \cite{q5} and  the amended hopping integral $t_{i}^{'}$  is 
\begin{eqnarray}\label{xt}
t_{i}^{'}=t_{0}e^{-3.37\left(|d^{'}_{i}|/a-1\right)}, \qquad i=1,2,3.
\end{eqnarray}

In the forthcoming analysis, the laser field of frequency $\omega$ is depicted by the potential vector in the electric dipole approximation \cite{Dipole} as
\begin{eqnarray}\label{xtz7}
\textbf{A}=\left(0,A\cos{\omega t}\right)\end{eqnarray}
and then the corresponding electric field is
\begin{eqnarray}\label{xt7}
\textbf{F}=-\dfrac{\partial}{\partial t}\textbf{A}=\left(0,F\sin{\omega t}\right)
\end{eqnarray} 
where $F=\omega A$ is the electric field amplitude. After substitution in (\ref{H1}) we obtain 
\begin{align}\label{H2}
\mathcal{H}&=\begin{pmatrix}
0&v_{x}p_{x}-i{v_{y}} \left(p_{y}-\dfrac{Fe}{\omega}\cos{\omega t}\right) \\
v_{x}p_{x} + iv_{y} \left(p_{y}-\dfrac{Fe}{\omega}\cos{\omega t}\right)   & 0
\end{pmatrix}.
\end{align}
In the presence of the  electromagnetic wave (laser field), the electron wavefunction and the band structure of dressing graphene are given into account within the framework of the {Floquet} theory of periodically systems \cite{a1,a3,a4,ai}. Based on this approximation, we write  
\begin{align}\label{gg}
\Psi_{j}(x,y,t)=e^{-i\frac{\mathcal{\varepsilon}t}{ \hbar}}\Phi_{j}(x,y,t)	= \left[\Psi_{A,j}(x,y,t),\Psi_{B,j}(x,y,t)\right]^{T}
\end{align}
where $\mathcal{\varepsilon}$ is the Floquet quasi-energy (energy of dressed electron) and $\Phi_{j}(x,y,t)$  is periodically function with
respect to $\omega t$ of period $2 \pi$, i.e. $\Phi_{j}(x,y,t)=\Phi_{j}(x,y,t + 2\pi/\omega)$.

Taking into account of such periodicity, we can  develop the full solution of system in \textbf{region 2} ($0\leq x\leq L$) in Fourier series 
\cite{Xu4,Xu5,f1}
\begin{align}\label{gg2}
\Psi_{2}(x,y,t)=e^{-i\frac{\mathcal{\varepsilon}t}{\hbar}}\Phi_{2}(x,y)  e^{-i\frac{v_{y}Fe}{\hbar\omega^{2}}\sin{\omega t}}
\end{align}
and by involving the generating function associated to the Bessel function $J_{n}$ of first kind we   write 
\begin{align}\label{g1}
\Psi_{2}(x,y,t)=\Phi_{2}(x,y) \sum^{
+\infty}_{n=-\infty} J_{n}\left(\frac{v_{y}Fe}{\hbar\omega^{2}}\right)e^{-\frac{i}{\hbar}\left({\mathcal{\varepsilon}}+n\hbar\omega\right)t}.
\end{align}
Since the laser potential barrier can be exchanged with the electron energies in units of $\hbar\omega$, then we construct  the eigenspinors (\ref{g1}) as a linear combination of those at all energies $\mathcal{\varepsilon}+m\hbar\omega$ ($m=0,\pm 1, \pm 2, \cdots$). Because the Hamiltonian is translational invariant along $y$-direction,  (\ref{g1}) becomes
\begin{align}\label{g11}
	\Psi_{2}(x,y,t)=e^{i k_{y}y}\sum^{
		+\infty}_{m,n=-\infty}\Phi^{m}_{2}(x)J_{n-m}{\left(\frac{ v_{y}Fe}{\hbar\omega^{2}}\right)e^{-\frac{i}{\hbar}\left({\mathcal{\varepsilon}}+n\hbar\omega\right)t}}
\end{align}
with the spinor $\Phi^{m}_{2}(x)$ contains a new index $m$ matching to the Floquet side band energy states, which will be determined  at  later stage. The eigenvalue equation gives
\begin{align}\label{hy}
\begin{pmatrix}
-i\left(\mathcal{\varepsilon}+m\hbar\omega\right)	&  \hbar\left(v_{x}\partial_{x}+v_{y}k_{y}-m\omega\right) \\
\hbar\left(v_{x}\partial_{x}-v_{y}k_{y}+m\omega\right)&  -i\left(\mathcal{\varepsilon}+m\hbar\omega\right)
\end{pmatrix}\begin{pmatrix}
 \Phi^{m}_{A,2}(x)\\
 \Phi^{m}_{B,2}(x)
\end{pmatrix}=\begin{pmatrix} 0\\0
\end{pmatrix}
\end{align}
which gives rise to 
the eigenvalues 
\begin{align}\label{Aqf}
	\mathcal{\varepsilon}+m\hbar\omega&=s_m\hbar\sqrt{{{v^{2}_{x}(q^{m}_{x})^{2}+\left({{v_{y}}k_{y}-m\omega}\right)^{2}}}}
\end{align}
where $s_m=\text{sgn}\left(\mathcal{\varepsilon}+m {\hbar}\omega\right)$ corresponds to the conduction and valence bands. The wave vector  $q^{m}_{x}$ can be obtained from (\ref{Aqf})
\begin{align}\label{Ahq}
	q^{m}_{x}=s_m\sqrt{\left(\dfrac{\mathcal{\varepsilon}+m\hbar\omega}{\hbar v_{x}}\right)^{2}-\left(\dfrac{{v_{y}}k_{y}-m\omega}{v_{x}}\right)^{2}}.
\end{align}
At this level, we have some comments in order. Indeed, (\ref{Ahq}) depends on different parameters including the energy, $y$-component of the wave vector,  frequency and   effective Fermi velocities. Moreover, (\ref{Ahq}) is modified (dressed) by the linearly polarized electromagnetic wave (laser field) {in different way   compared to  the  time dependent scalar potential \cite{X8,ln,lp,lo}}. Additionally, we recover the results obtained in \cite{f1} as limiting case  $S=0\%$. As for  the  eigenspinors we find
\begin{align}\nonumber
	\Phi^{m}_{2}(x,y)&=e^{ik_{y}y}\left[C_{m}\begin{pmatrix}
		1\\
		s_m e^{i\varphi_{m}}
	\end{pmatrix}e^{iq^{m}_{x}x}+D_{m}\begin{pmatrix}
		1\\
		-s_m e^{-i\varphi_{m}}
	\end{pmatrix}e^{-iq^{m}_{x}x}\right].
\end{align}
Ultimately, combining all to obtain the solutions in \textbf{region 2} ($0\leq x\leq L$) in the presence of the laser field as 
\begin{align}\nonumber
	\Psi_{2}(x,y,t)&=e^{ik_{y}y}\sum^{
		+\infty}_{m,n=-\infty} \left[C_{m}\begin{pmatrix}
		1\\
	{\color{black}s_m  e^{i\varphi_{m}}}
	\end{pmatrix}e^{iq^{m}_{x}x}+D_{m} \begin{pmatrix}
	1\\
	-s_m e^{-i\varphi_{m}}
\end{pmatrix}e^{-iq^{m}_{x}x}\right]J_{n-m}{\left(\frac{v_{y}Fe}{\hbar\omega^{2}}\right)}e^{-\frac{i}{{\color{black}\hbar}}\left(\mathcal{\varepsilon}+n{\hbar}\omega\right)t}
\end{align}
where
$C_{m}$, $D_{m}$ are two constants and
 we have set
 \begin{align}
 \Gamma_{m}= 
\dfrac{{\hbar}\left[{v_{x}} q^{m}_{x}+i\left( {{v_{y}}}k_{y}-m\omega\right)\right]}{\mathcal{\varepsilon}+m {\hbar}\omega}={s_m}e^{i\varphi_{m}}, \qquad \varphi_{m}=\tan^{-1}\left(\dfrac{ {{v_{y}}}k_{y}-m\omega}{{v_{x}} q^{m}_{x}}\right).
 \end{align}

For \textbf{region 1} $(x<0)$ in the absence of strain $S=0\%$ and by neglecting the effect of the laser field under the consideration of the electric dipole approximation, the solutions of incident wave 
 of energy $\mathcal{\varepsilon}$  are given by 
\begin{align}
	\Psi_{\textbf{in}}(x,y,t)&=A_{0}\begin{pmatrix}
		1\\
		\dfrac{{\hbar v_{F}}\left[k^{0}_{x}+ik_{y}\right]}{\mathcal{\varepsilon}}
	\end{pmatrix}e^{ik^{0}_{x}x+ik_{y}y}e^{-i\frac{\mathcal{\varepsilon}t}{{\hbar}}}
\end{align}
while those of energies $\mathcal{\varepsilon}+m\hbar\omega$
for the reflected waves take the forms
\begin{align}
	\Psi_{\textbf{re}}(x,y,t)&=e^{ik_{y}y}\sum^{
		+\infty}_{m,n=-\infty} r_{m} \begin{pmatrix}
		1\\
		\dfrac{{\hbar v_{F}}\left[-k^{m}_{x}+ik_{y}\right]}{\mathcal{\varepsilon}+m{\hbar}\omega}
	\end{pmatrix}e^{-ik^{m}_{x}x}{e^{-\frac{i}{\hbar}\left(\mathcal{\varepsilon}+m\hbar\omega\right)t}}
\end{align}
associated to the wave vector  \begin{align}\label{Ar1}
k^{m}_{x}=s'_m\sqrt{\left(\dfrac{\mathcal{\varepsilon}+m\hbar\omega}{\hbar v_{F}}\right)^{2}-k^{2}_{y}}
\end{align}
Consequently, the eigenspinors in  \textbf{region 1} $(x<0)$  are
\begin{align}\nonumber
\Psi_{1}(x,y,t)&=e^{ik_{y}y}\left[A_{0} \begin{pmatrix}
1\\
s_0e^{i\theta_{0}}
\end{pmatrix}e^{ik^{0}_{x}x}e^{-i\frac{\mathcal{\varepsilon}t}{\hbar}}
+\sum^{+\infty}_{m=-\infty} r_{m}  \begin{pmatrix}
	1\\
	{-s'_me^{-i\theta_{m}}}
\end{pmatrix}e^{-ik^{m}_{x}x}e^{-\frac{i}{{\hbar}}\left(\mathcal{\varepsilon}+m{\hbar}\omega\right)t}\right]
\end{align}
where $s_0=\text{sgn}(\mathcal{\varepsilon})$,
$A_{0}$ and $r_{m}$ are the amplitudes of the wave incident and  reflection, respectively. Here
we have
\begin{align}\label{at1}
\Lambda_{m}=\dfrac{{\hbar v_{F}}\left[k^{m}_{x}+ik_{y}\right]}{\mathcal{\varepsilon}+m {\hbar}\omega}=s'_m e^{i\theta_{m}},
\qquad 
\theta_{m}= \tan^{-1}\left(\dfrac{k_{y}}{k^{m}_{x}}\right)
\end{align}
Regarding  \textbf{region 3} $(x>L)$, we immediately derive the eigenspinors $\Psi_{3}(x,y,t)$ for transmitted fermions as done in \textbf{region 1}
\begin{align}\nonumber
	\Psi_{3}(x,y,t)&=e^{ik_{y}y}\sum^{
		+\infty}_{m=-\infty}\left[t_{m} \begin{pmatrix}
	1\\
	s'_me^{i\theta_{m}}
\end{pmatrix}e^{ik^{m}_{x}x}+ F_{m} \begin{pmatrix}
1\\
-s'_me^{-i\theta_{m}}
\end{pmatrix}e^{-ik^{m}_{x}x}\right]e^{-\frac{i}{{\hbar}}\left(\mathcal{\varepsilon}+m{\hbar}\omega\right)t}
\end{align}
with $t_{m}$ is the amplitude of transmission and $F_{m}$ is the null vector.

It is clearly seen that from the results obtained above the solutions of \textbf{region 2} show strong dependency on the strain effect along armchair and zigzag directions, which is not the case for the solutions of \textbf{region 1} and \textbf{3}. In the next section, we will study their impact on the transmission probability trough laser barrier.

\section{Transmission probability}
The transmission coefficient can be obtained by considering the orthogonality of $\big\{e^{i n\omega t}\big\}$ and the continuity {\color{black}condition  of wave functions} at boundaries ($x=0$, $x=L$) in the context of the transfer matrix approach \cite{fb1,fa1}, to end up with
a set of equations for $A_{n}$, $r_{n}$, $C_{n}$, $D_{n}$, $t_{n}$, $F_{n}$. Thus at interface $x = 0$, one finds \begin{align}\label{at}
&
A_{n}+r_{n}=\sum^{+\infty}_{m=-\infty}\left[C_{m}+D_{m}\right]J_{n-m} \left(\frac{v_{y}Fe}{\hbar\omega^{2}}\right)\\
&
A_{n}\Lambda_{n}-{r_{n}}{\frac{1}{\Lambda_{n}}}=\sum^{+\infty}_{m=-\infty}\left[C_{m}\Gamma_{m}-{D_{m}}{\frac{1}{\Gamma_{m}}}\right]J_{n-m} {\left(\frac{ v_{y}Fe}{\hbar\omega^{2}}\right)}
\end{align}
with $A_{n}=\delta_{n,0}$.  At $x=L$, we have
\begin{align}
&
t_{n}e^{ik^{n}_{x}L}+F_{n}e^{-ik^{n}_{x}L}=\sum^{+\infty}_{m=-\infty}\left[C_{m}e^{iq^{m}_{x}L}+D_{m}e^{-iq^{m}_{x}L}\right]J_{n-m} {\left(\frac{ v_{y}Fe}{\hbar\omega^{2}}\right)}\\
\label{atb}&t_{n}\Lambda_{n}e^{ik^{n}_{x}L}-{F_{n}}{\frac{1}{\Lambda_{n}}}e^{-ik^{n}_{x}L}=\sum^{+\infty}_{m=-\infty}\left[C_{m}\Gamma_{m}e^{iq^{m}_{x}L}-{D_{m}}{\frac{1}{\Gamma_{m}}}e^{-iq^{m}_{x}L}\right]J_{n-m} {\left(\frac{v_{y}Fe}{\hbar\omega^{2}}\right)}
\end{align}
These boundary conditions can be displayed in transfer matrix formalism as 
\begin{align}\label{ar1}
	\begin{pmatrix}
		\mathbb{G}_{1} \\
		\mathbb{G}^{'}_{1}
	\end{pmatrix}=&\begin{pmatrix}
		\mathbb{N}_{11}	&\mathbb{N}_{12}\\
		\mathbb{N}_{21}& \mathbb{N}_{22}
	\end{pmatrix}	\begin{pmatrix}
		\mathbb{G}_{2} \\
		\mathbb{G}^{'}_{2}
	\end{pmatrix}=\mathbb{N}	\begin{pmatrix}
		\mathbb{G}_{2} \\
		\mathbb{G}^{'}_{2}
	\end{pmatrix}
\end{align}
and $\mathbb{N}=\mathbb{N}(1,2) \cdot\mathbb{N}(2,3)$ is the total transfer matrix with
\begin{align}\label{Na}
	\mathbb{N}(1,2)&=\begin{pmatrix}
		\mathbb{I}&	\mathbb{I}\\
		\mathbb{M}^{+}_{0}& \mathbb{M}^{-}_{0}
	\end{pmatrix}^{-1}\begin{pmatrix}
		\mathbb{K}_{0}&\mathbb{K}_{0}\\
		\mathbb{P}^{+}_{0}& \mathbb{P}^{-}_{0}
	\end{pmatrix}\\\label{Na1}\mathbb{N}(2,3)&=\begin{pmatrix}
		\mathbb{K}^{+}_{L}&\mathbb{K}^{-}_{L}\\
		\mathbb{P}^{+}_{L}& \mathbb{P}^{-}_{L}
	\end{pmatrix}^{-1}\begin{pmatrix}
		\mathbb{I}&	\mathbb{I}\\
		\mathbb{M}^{+}_{0}& \mathbb{M}^{-}_{0}
	\end{pmatrix}\begin{pmatrix}
		\mathbb{Q}^{+}&	\mathbb{O}\\
		\mathbb{O}& \mathbb{Q}^{-}
	\end{pmatrix}
\end{align}
where we have defined the quantities
\begin{align}\label{ma}
(\mathbb{M}^{\pm}_{0})_{n,m}&=\pm(\Lambda_{n})^{\pm1}\delta_{n,m}, \qquad\qquad\qquad\qquad\quad (\mathbb{P}^{\pm}_{0})_{n,m}=\pm(\Gamma_{m})^{\pm1}J_{n-m}{\left(\frac{v_{y}Fe}{\hbar\omega^{2}}\right)}\\ (\mathbb{K}_{0})_{n,m}&=J_{n-m} {\left(\frac{ v_{y}Fe}{\hbar\omega^{2}}\right)},  \qquad\qquad\qquad\qquad\quad(\mathbb{K}^{\pm}_{L})_{n,m}=e^{\pm iq^{m}_{x}L}J_{n-m} {\left(\frac{ v_{y}Fe}{\hbar\omega^{2}}\right)}\\ (\mathbb{P}^{\pm}_{L})_{n,m}&=\pm(\Gamma_{m})^{\pm1}e^{\pm iq^{m}_{x}L}J_{n-m} {\left(\frac{ v_{y}Fe}{\hbar\omega^{2}}\right)},\qquad (\mathbb{Q}^{\pm})_{n,m}=e^{\pm ik^{n}_{x}L}\delta_{n,m}
\end{align}
and  the coefficients
\begin{align}\label{mz1}
\mathbb{G}_{1}&=\big\{A_{m}\}, \qquad \mathbb{G}^{'}_{1}=\big\{r_{m}\}, \qquad \mathbb{G}_{2}=\big\{t_{m}\}, \qquad  \mathbb{G}^{'}_{2}=\big\{F_{m}\}=0 
\end{align} 
with  $\mathbb{O}$ and $\mathbb{I}$ indicate the unit and null matrices. From the above process and (\ref{ar1}), we get the relation 
\begin{align}\label{ma1}
\mathbb{G}_{2}=(\mathbb{N}_{11})^{-1}\mathbb{G}_{1} 	
\end{align}
We {\color{red}clearly seen} that the linear series of equations (\ref{at}-\ref{atb}) have an infinite number of unknowns ($m, n$ goes from $-\infty$ to $+\infty$) and can be truncated to consider a limited number of terms starting from $-N$ up to $N$ \cite{lp,f2}. Then, (\ref{ma1}) becomes 
\begin{align}
t_{-N+k}=\mathbb{N}^{'}[k+1,N+1]
\end{align}
where $k=0,1,2,3,\cdots, 2N$ and $\mathbb{N}^{'}$ is the inverse matrix $\mathbb{N}_{11}^{-1}$. 

To derive the transmission coefficient, we introduce the current density 
\begin{align}\label{ma4a}
	J=e v_{F}\Psi^{\dagger}(x,y,t)\sigma_{x}\Psi(x,y,t)
\end{align}
where $\sigma_{x}$ is the Pauli matrix and $\Psi(x,y,t)$ are the solutions of the energy spectrum. Thereby, (\ref{ma4a}) gives the incident $J_{\text{in},0}$ and transmitted $J_{\text{tr},m}$ components of the current densities 
\begin{align}\label{ma5a} 
	J_{\text{in},0}&={2s_0e v_{F} \dfrac{A^{2}_{0}k^{0}_{x}}{\sqrt{(k^{0}_{x})^{2}+k^{2}_{y}}}}\\\label{ma5b}J_{\text{tr},m}&={2s'_me v_{F} \dfrac{t^{2}_{m}k^{m}_{x}}{\sqrt{(k^{m}_{x})^{2}+k^{2}_{y}}}}
\end{align}
Ultimately, the transmission coefficient for the $m^{th}$ sideband is given by
\begin{align}\label{ma3}
	T_{m}=\left|\dfrac{J_{\text{tr},m}}{J_{\text{in},0}}\right|=\lambda_{m}\left|\dfrac{t_{m}}{A_{0}}\right|^{2}
\end{align}
where the parameter $\lambda_{m}$ is defined by 
\begin{align}\label{ma8}
\lambda_{m}={\dfrac{s'_mk^{m}_{x}}{s_0k^{0}_{x}}\frac{\sqrt{(k^{0}_{x})^{2}+k^{2}_{y}}}{\sqrt{(k^{m}_{x})^{2}+k^{2}_{y}}}}={\dfrac{s'_m}{s_0}}\dfrac{\cos{\theta_{m}}}{\cos{\theta_{0}}}
\end{align}
Inversely of the barrier static and like the oscillating barrier the total transmission coefficient of laser barrier structure is given by the sum over all modes $m$
\begin{align}
	T_{c}=\sum_{m}T_{m}
\end{align}
Due to numerical difficulties, we limit ourselves to three transmission channels, which are the  central and two first sidebands $m=-1,0,1$
\begin{align}\label{me}
	t_{-1}=\mathbb{N}^{'}[1,2], \qquad 	t_{0}=\mathbb{N}^{'}[2,2],\qquad t_{1}=\mathbb{N}^{'}[3,2]
\end{align}
{Consequently, we will analyze numerically these results under suitable conditions of the barrier width $L$, laser field amplitude $F$, frequency $\omega$ and strain amplitude $S$.}

\section{Results and discussions}
To underline the effect of strain along armchair and zigzag directions at normal incidence $\theta_{0}=0^{\circ}$  on the transmission probability, we present in {\color{blue}\textbf{Figure}} \ref{figB4}
{the central band $T_{0}$ and the first sidebands $T_{\pm 1}$} {
as function of the barrier width $L$ for $\mathcal{\varepsilon}=90$ \text{meV}, $\omega=7\times10^{12}$ \text{Hz}, $F=0.05$ \text{V/nm} with three values of the  strain magnitude $S=0\%$, $S=6.5\%, 20\%$ (A), $S=6.5\%, 20\%$ (Z). As far as $S=0\%$ is concerned see {\color{blue}\textbf{Figure}} \ref{figB4}\textbf{\color{blue}{(a)}}, the transmission through the central band predominates over all the sidebands and decreases for small values of the barrier width but shows a damped oscillatory behavior for its large values, {i.e. $L\leq650$ \text{nm} and after that it slowly increases}. This may be due to the occurrence of the Bessel function in the expression of transmission {in similar way to  an oscillating barrier in time \cite{X8}}. On the other hand, we clearly see that the transmission for the sidebands increases from zero and oscillates with a damped amplitude. It is shown that at normal incidence, the laser assisted transmission for the photon absorption $T_{+1}$ and emission $T_{-1}$ are identical in the interval $0<L<160$ \text{nm} but not identical from $L>160$ \text{nm}, contrary to the oscillating barrier where  $T_{+1}$ and $T_{-1}$ coincide whatever the values taken by $L$ {as has been reported in \cite{X8,lp,lo}}. Note that the transmission considered along armchair and zigzag directions is showing different behaviors and it increases progressively by oscillating under the increase of the  barrier width $L$ and strain magnitude $S$. Indeed, we observe that the amplitude of the central band transmission which located in the interval $250$ \text{nm} $<L<650$ \text{nm} when $S=0\%$ becomes very small if the deformation is applied along armchair direction with $S=6.5\%$ and it disappears for $S=20\%$ as shown in {\color{blue}\textbf{Figure}} \ref{figB4}\textbf{\color{blue}{(b)}}. Also, it is seen that the obvious change on oscillations of sidebands transmission, which is  a manifestation of a $S$ non-null. {\color{blue}\textbf{Figure}} \ref{figB4}\textbf{\color{blue}{(c)}} shows that the strain along zigzag direction produces a remarkable influence on $T_{c}$. Indeed,  as long as $S$ increases, the amplitude  and period of the transmission get modified as well as the number of peaks increases. It is clearly seen  that $S$ affects all transmission channels and then can be used as a key component to control tunneling properties of our system. 
\begin{figure}[H]
	\centering
	\subfloat[]{
		\includegraphics[width=0.327\linewidth, height=0.24\textheight]{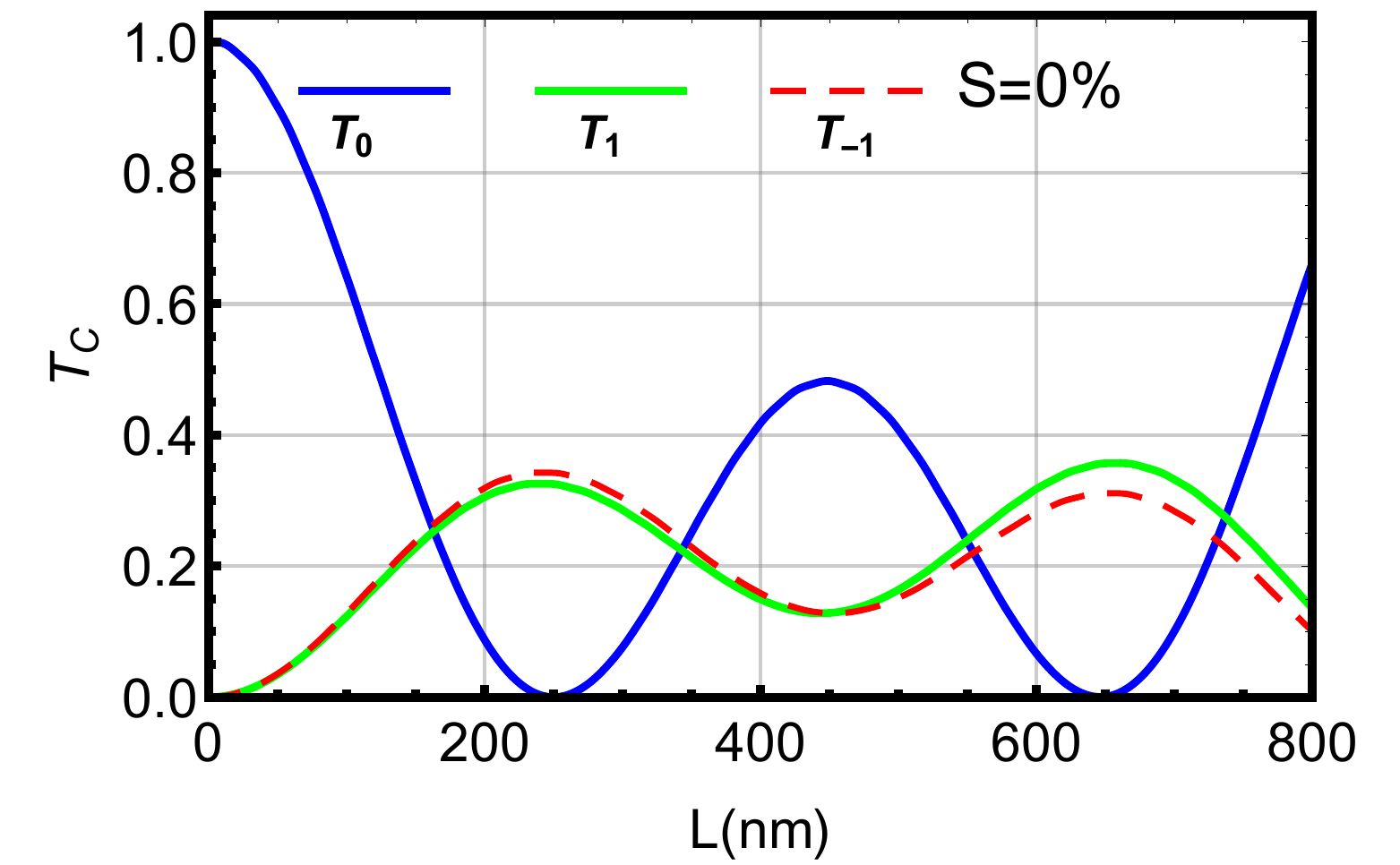}\label{gK}}
	\subfloat[]{
		\includegraphics[width=0.327\linewidth, height=0.24\textheight]{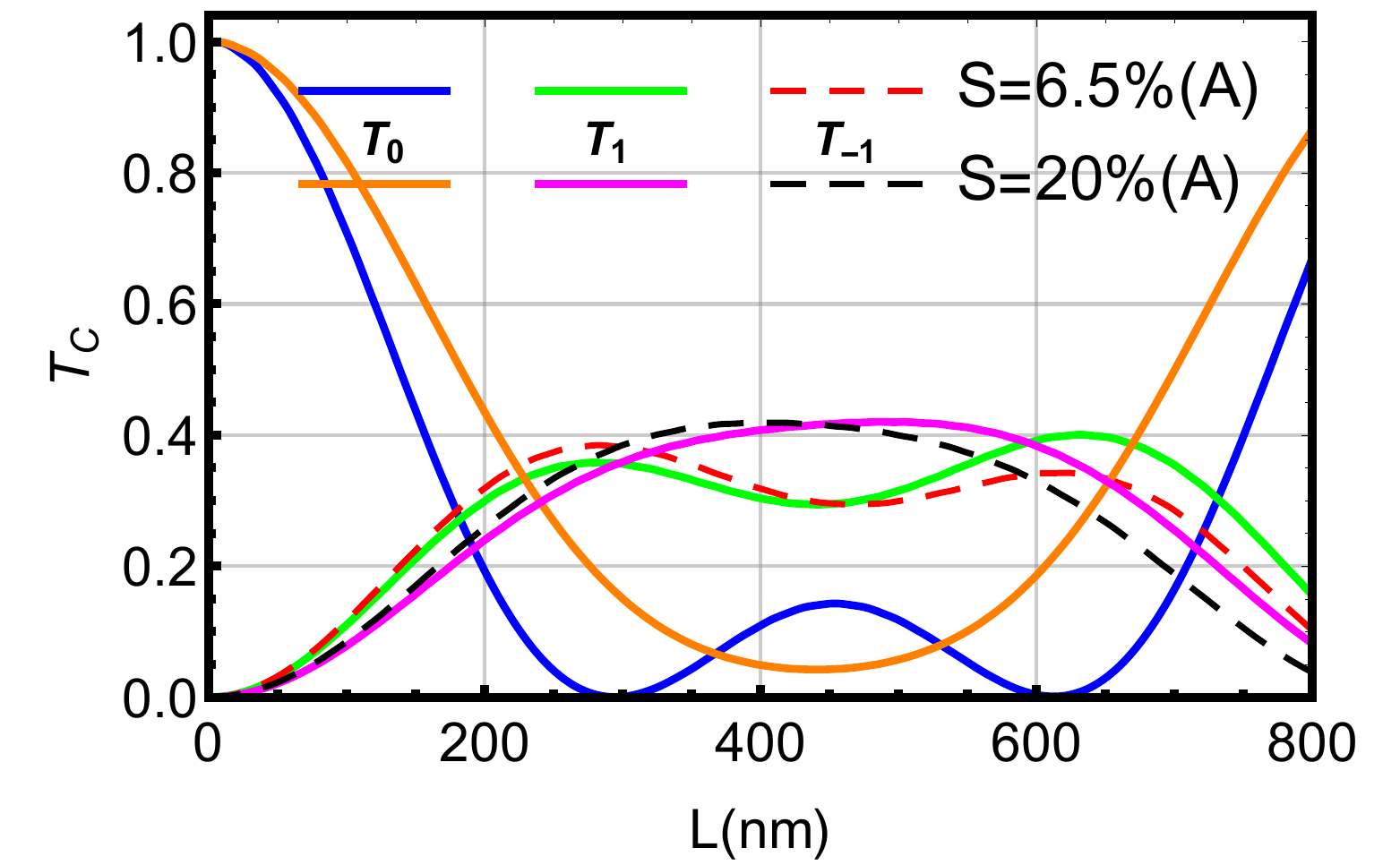}\label{gK1}}
	\subfloat[]{
		\includegraphics[width=0.327\linewidth, height=0.24\textheight]{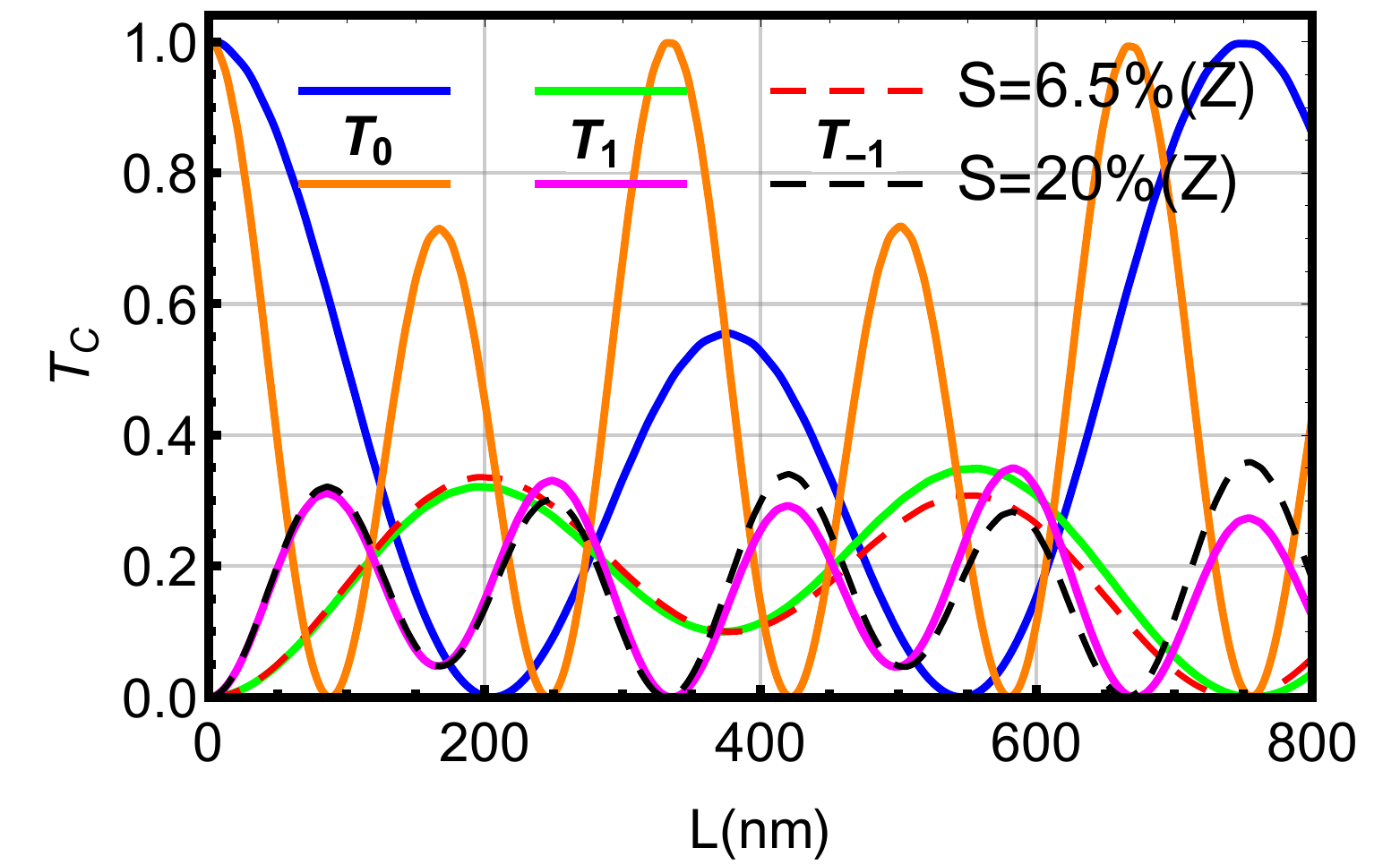}\label{gK2}}	
	\caption{\sf{(color online) Transmission probabilities {of the central band $T_{0}$ and first sidebands $T_{\pm 1}$} 
			as function of the barrier width $L$ at normal incidence $\theta_{0}=0^{\circ}$ for  $\mathcal{\varepsilon}=90$ \text{meV}, $\omega=7\times10^{12}$ \text{Hz}, $F=0.05$ \text{V/nm}.  \textbf{\color{blue}{(a)}}\color{black}{:} Without strain $S=0\%$. \textbf{\color{blue}{(b)}}\color{black}{:} Effect of armchair strain direction for $S=6.5\%, 20\%$. \textbf{\color{blue}{(c)}}\color{black}{:} Effect of zigzag strain direction for $S=6.5\%, 20\%$.}}\label{figB4}
\end{figure}

{\color{blue}\textbf{Figure}} \ref{figB41} presents the transmission probabilities {the central band $T_{0}$ and first sidebands $T_{\pm 1}$}  
as function of the barrier width $L$ for  opposite incident angles $\theta_{0}=\pm 20^{\circ}$ with different strain magnitudes and $\mathcal{\varepsilon}=90$ \text{meV}, $\omega=7\times10^{12}$ \text{Hz}, $F=0.05$ \text{V/nm}. It is clearly shown that the transmission of the central band and the other two nearest sidebands shows 
{an} oscillatory behavior close to that obtained for normal incidence except that the peaks of the transmission are shifted to the right for $\theta_{0}=-20^{\circ}$ and its number decreases but increases for $\theta_{0}=20^{\circ}$. {It is clearly seen that when $\theta_{0}=-20^{\circ}$ and for all cases, the transmission  of the central band displays a damped oscillatory nature with growing the values of $L$.}  Moreover, we notice that the oscillation frequency becomes larger for the positive value of $\theta_{0}$ compared to  the negative one. {These characteristics are somewhat similar to those obtained when the transmission is plotted versus the laser coupling parameter \cite{f1,Xu5}}. It is important to note that for $\theta_{0}=20^{\circ}$ the transmission exhibits 
the Fano type resonance almost similar to the tunneling through quantum wells in conventional semiconductor \cite{gt}. Therefore, we conclude that these effects depend on the sign of the incident angle $\theta_{0}$ and direction of the applied strain.
\begin{figure}[H]
	\centering
	\subfloat[]{
		\includegraphics[width=0.327\linewidth, height=0.24\textheight]{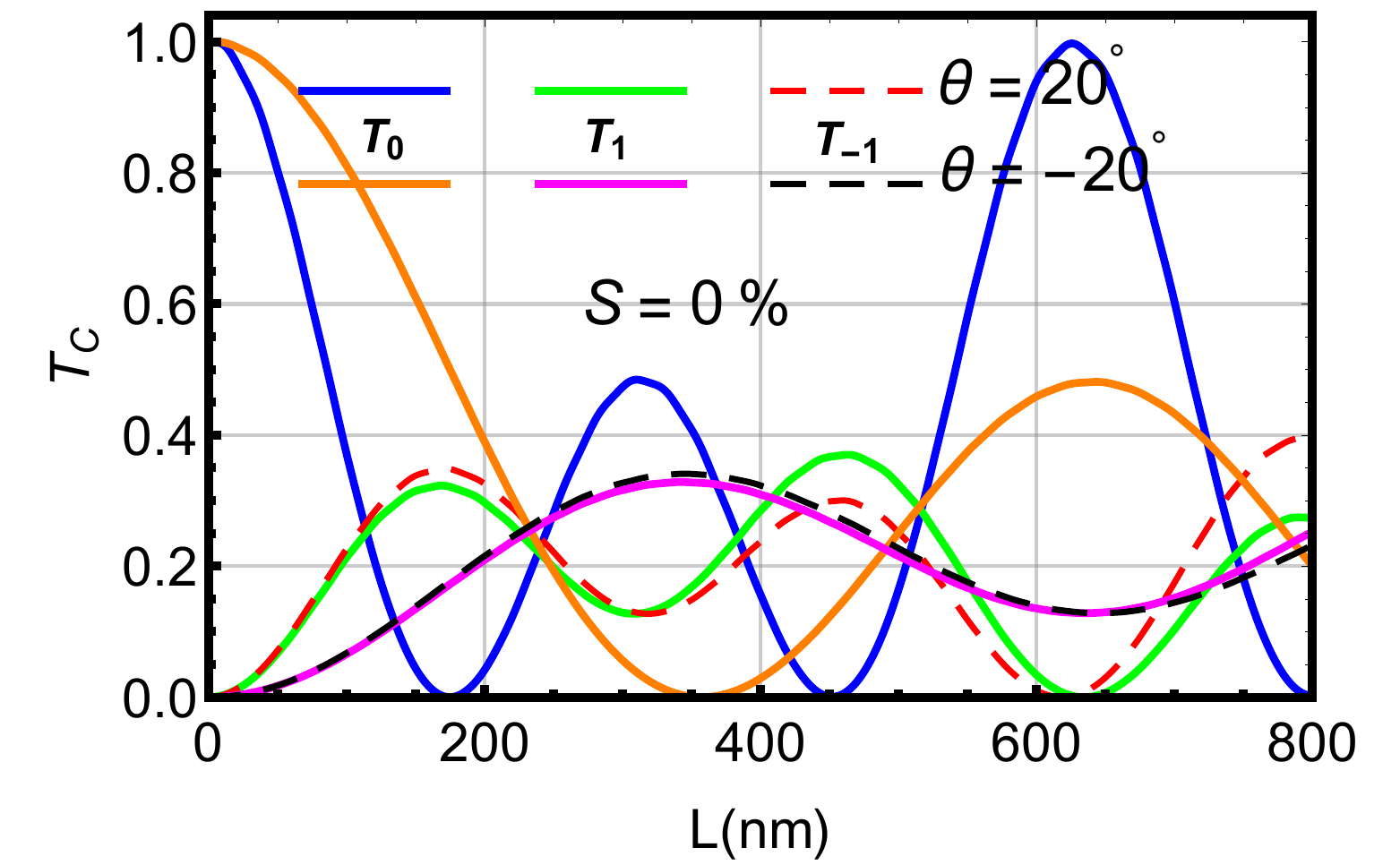}\label{gK3}}
	\subfloat[]{
		\includegraphics[width=0.327\linewidth, height=0.235\textheight]{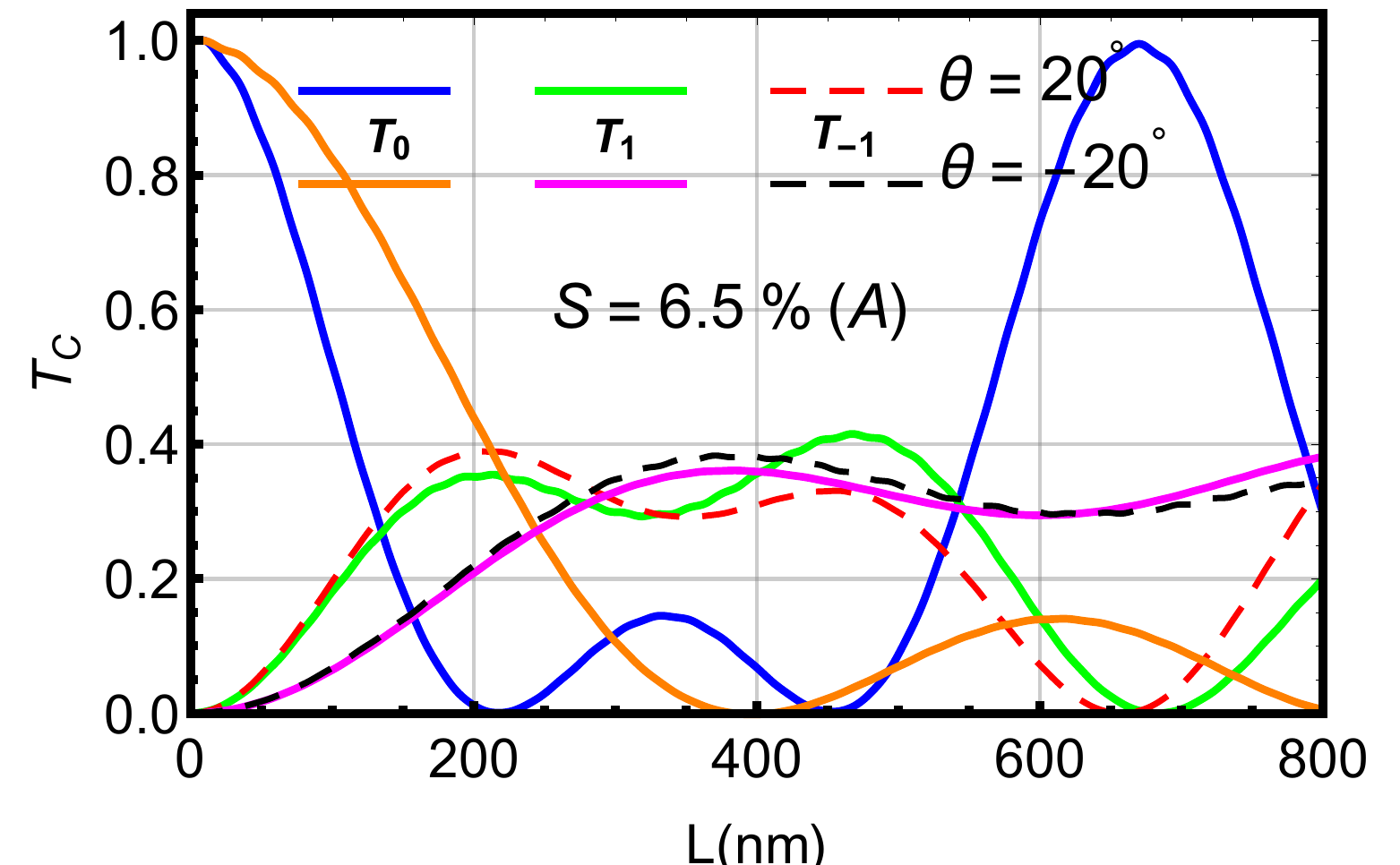}\label{gK4}}
	\subfloat[]{
		\includegraphics[width=0.327\linewidth, height=0.24\textheight]{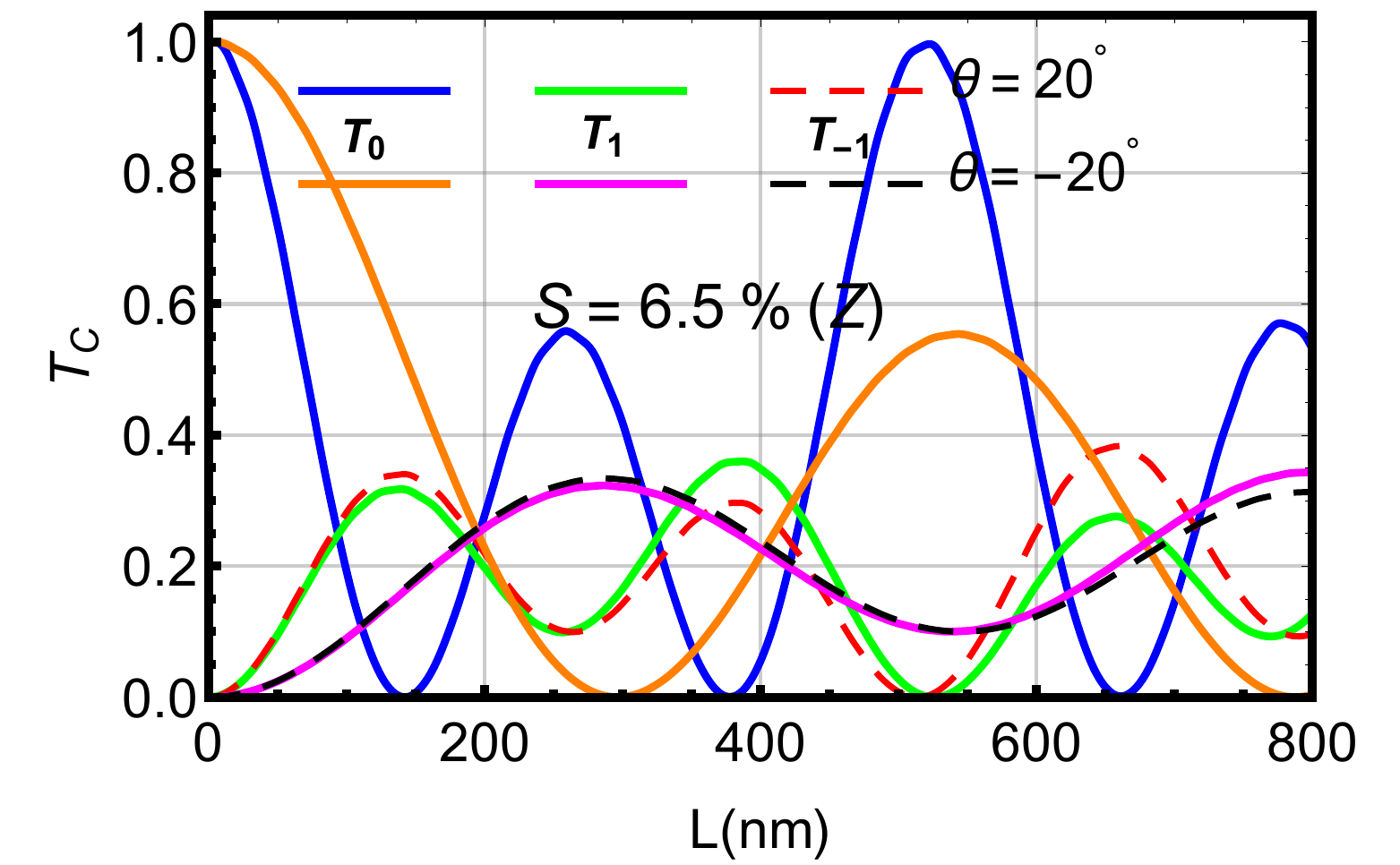}\label{gK5}}	
	\caption{\sf{(color online) Transmission probabilities {of the central band $T_{0}$ and first sidebands $T_{\pm 1}$} 
				as function of the barrier width $L$
	as function of the barrier width $L$  for opposite incident angles $\theta_{0}=\pm 20^{\circ}$ with $\mathcal{\varepsilon}=90$ \text{meV}, $\omega=7\times10^{12}$ \text{Hz}, $F=0.05$ \text{V/nm}.  \textbf{\color{blue}{(a)}}\color{black}{:} Without strain $S=0\%$. \textbf{\color{blue}{(b)}}\color{black}{:} Effect of armchair strain direction for $S=6.5\%$. \textbf{\color{blue}{(c)}}\color{black}{:} Effect of zigzag strain direction for $S=6.5\%$.}}\label{figB41}
\end{figure}

To illustrate the effect of the laser field amplitude $F$ along different strain magnitudes at normal incidence $\theta_{0}=0^{\circ}$ on the tunneling spectra of Dirac fermion in graphene through the laser barrier structure, we show in {\color{blue}\textbf{Figure}} $\ref{fiz}$ the transmission probabilities  {of the central band $T_{0}$ and first sidebands $T_{\pm 1}$} 
as function of the barrier width $L$ for $\mathcal{\varepsilon}=90$ \text{meV}, $\omega=7\times10^{12}$ and three values of the laser field amplitude $F=0.015$ \text{V/nm} (blue and orange lines), $F=0.03$ \text{V/nm} (green and magenta lines), $F=0.045$ \text{V/nm} (red and black lines) with $S=0\%$, $S=20\%$ (A), $S=20\%$ (Z). For $S=0\%$, we emphasis that the transmission through the central band shows an oscillatory behavior and decreases by increasing  $F$.   Particularly, for $F=0.045$ \text{V/nm} (red line), we observe the appearance of the Fano type resonance peaks \cite{gt,gt1}. On other hand, we notice that when $L$ is small (close to zero) the sideband transmission for the single photon absorption $T_{+1}$ is still null. By increasing  the barrier width $L$, $T_{+1}$ starts to oscillate and takes a characteristic form of the Gaussian shape exactly for $F=0.015$ \text{V/nm} (blue line) as well as its amplitude increases rapidly when $F$ increases. Note that the sideband transmission for the single photon emission $T_{-1}$ contains the same behavior of the sideband transmission for the photon absorption $T_{+1}$ except that it is reversed {and shifted to the left}. In {\color{blue}\textbf{Figures}} \ref{fiz}\textbf{\color{blue}{(a)}}, \ref{fiz}\textbf{\color{blue}{(b)}}, \ref{fiz}\textbf{\color{blue}{(c)}} when the strain is applied along armchair direction with $S=20\%$, we observe that it affects $T_{c}$ and causes some changes where the transmission displaces to the up for the central band $T_{0}$ and to the down for the first sidebands $T_{\pm 1}$ as long as the laser field amplitude $F$ increases. Further, one can see for $F=0.045$ \text{V/nm} (black line) the sharp oscillations disappears and $T_{c}$ takes a similar form to the other cases. However, we clearly seen that there is a admirable impact by switching the strain to the zigzag direction. Indeed, according to {\color{blue}\textbf{Figures}} \ref{fiz}\textbf{\color{blue}{(d)}}, \ref{fiz}\textbf{\color{blue}{(e)}}, \ref{fiz}\textbf{\color{blue}{(f)}}  one observes that the increase in $F$ is accompanied by the emergence of new oscillations. In addition they
show the same behaviors as in {\color{blue}\textbf{Figures}} \ref{fiz}\textbf{\color{blue}{(a)}}, \ref{fiz}\textbf{\color{blue}{(b)}}, \ref{fiz}\textbf{\color{blue}{(c)}} concerning the increase/decrease of the amplitude of sideband/central band transmission for large values of $F$. Furthermore, we observe that the amplitude of the transmission via photon emission $T_{-1}$ becomes larger than the process via photon absorption $T_{+1}$ in the interval $650$ \text{nm} $<L<800$ \text{nm}. We conclude that for the case of a laser irradiation, the transmission coefficient can be controlled by tuning the laser field amplitude $F$, strain amplitude $S$ and barrier width $L$.
\begin{figure}[H]
	\centering
	\subfloat[]{
		\includegraphics[width=0.326\linewidth, height=0.24\textheight]{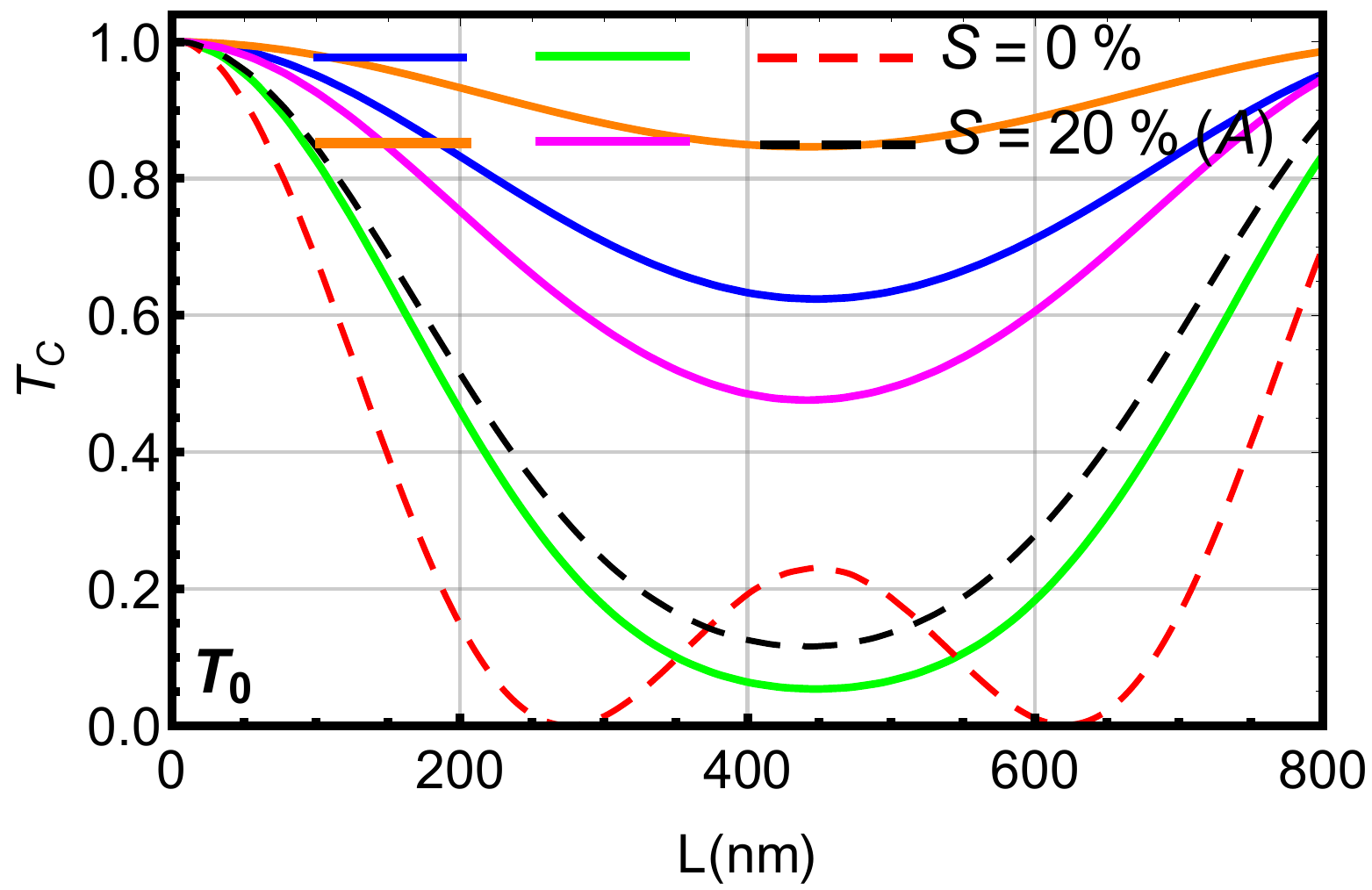}\label{akd}}
	\subfloat[]{
		\includegraphics[width=0.326\linewidth, height=0.24\textheight]{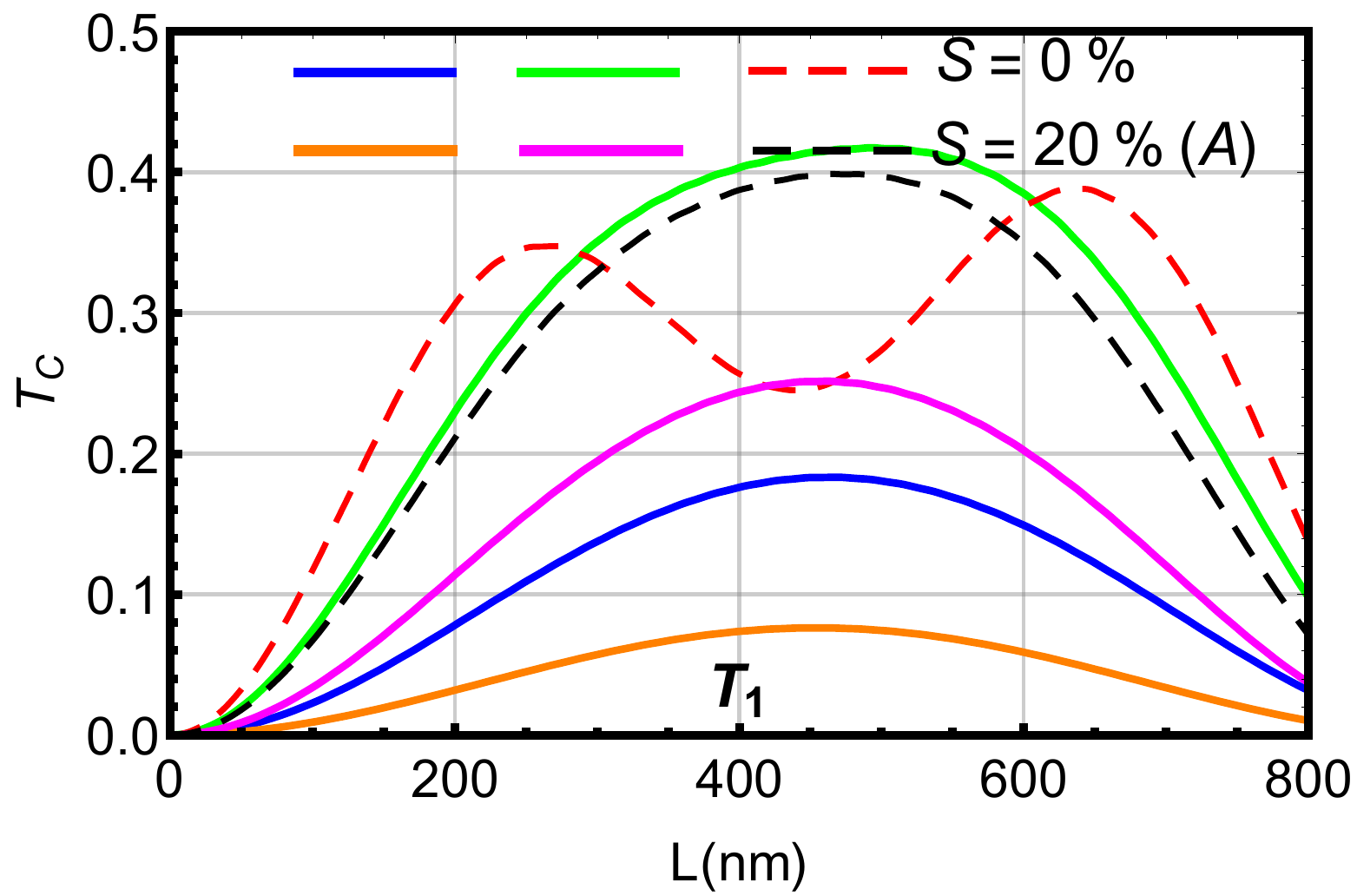}\label{kad1}}
	\subfloat[]{
		\includegraphics[width=0.326\linewidth, height=0.24\textheight]{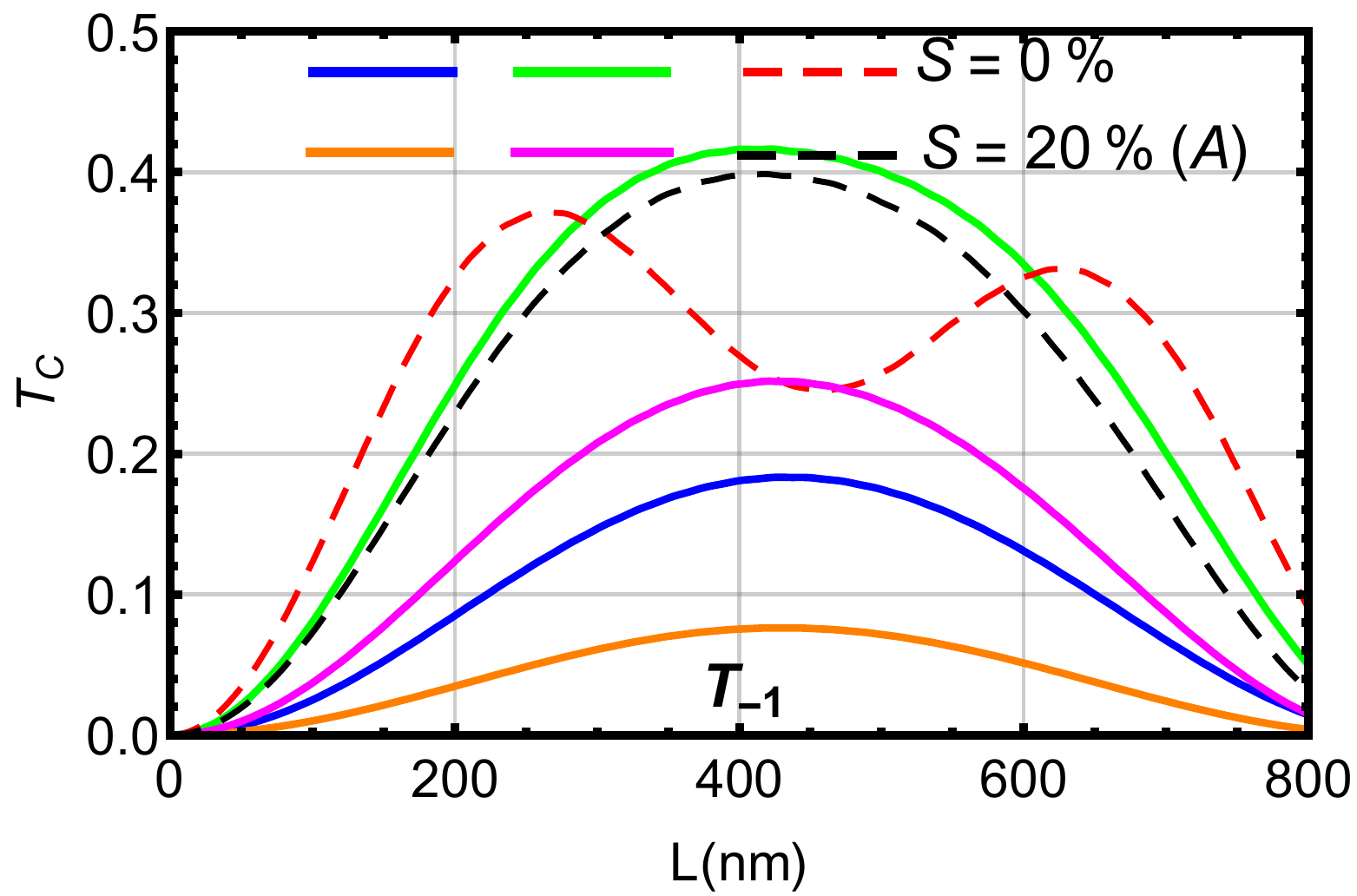}\label{kad2}}
	\\ 	\centering
	\subfloat[]{
		\includegraphics[width=0.326\linewidth, height=0.24\textheight]{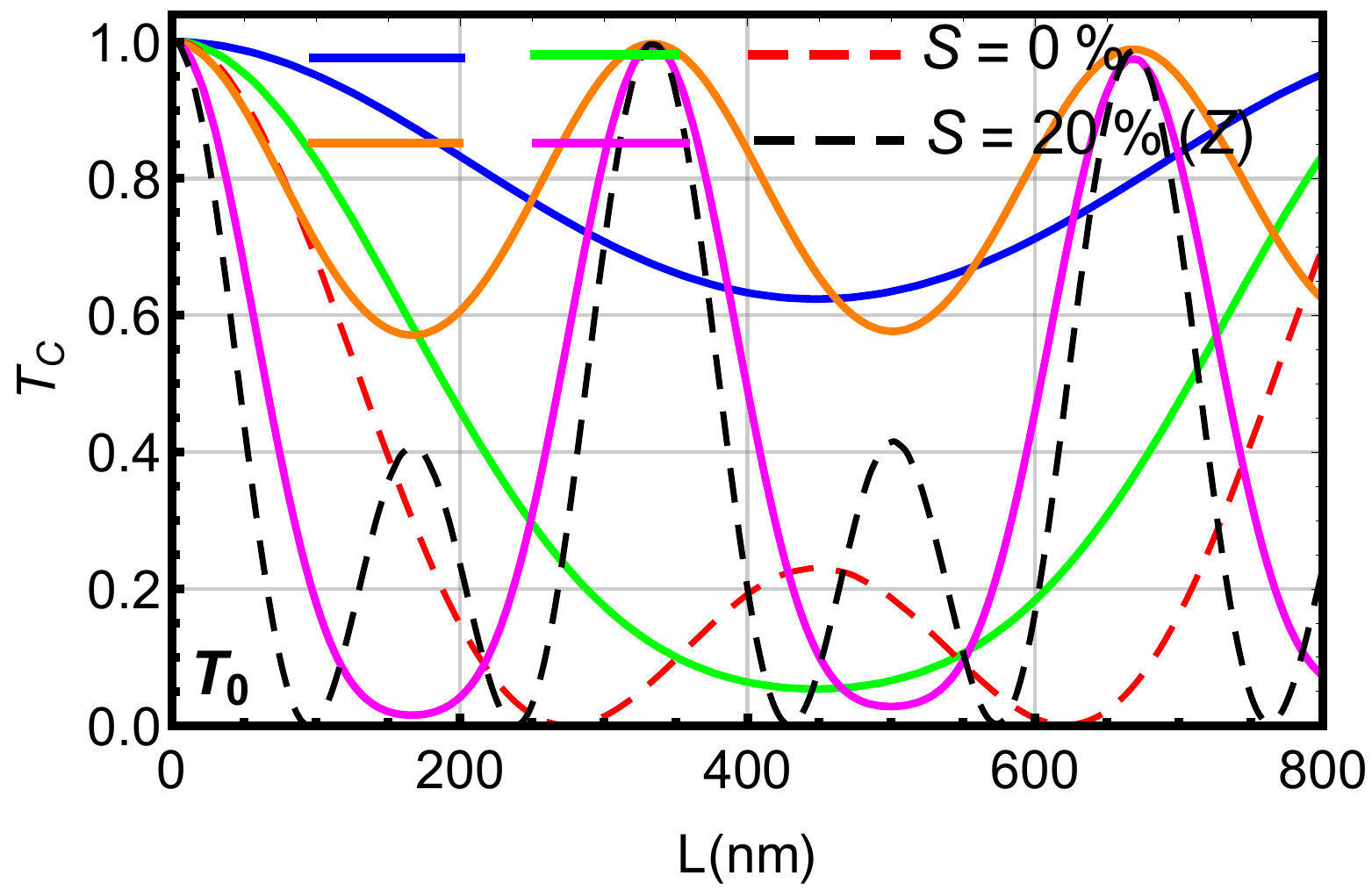}\label{kag}}
	\subfloat[]{
		\includegraphics[width=0.326\linewidth, height=0.24\textheight]{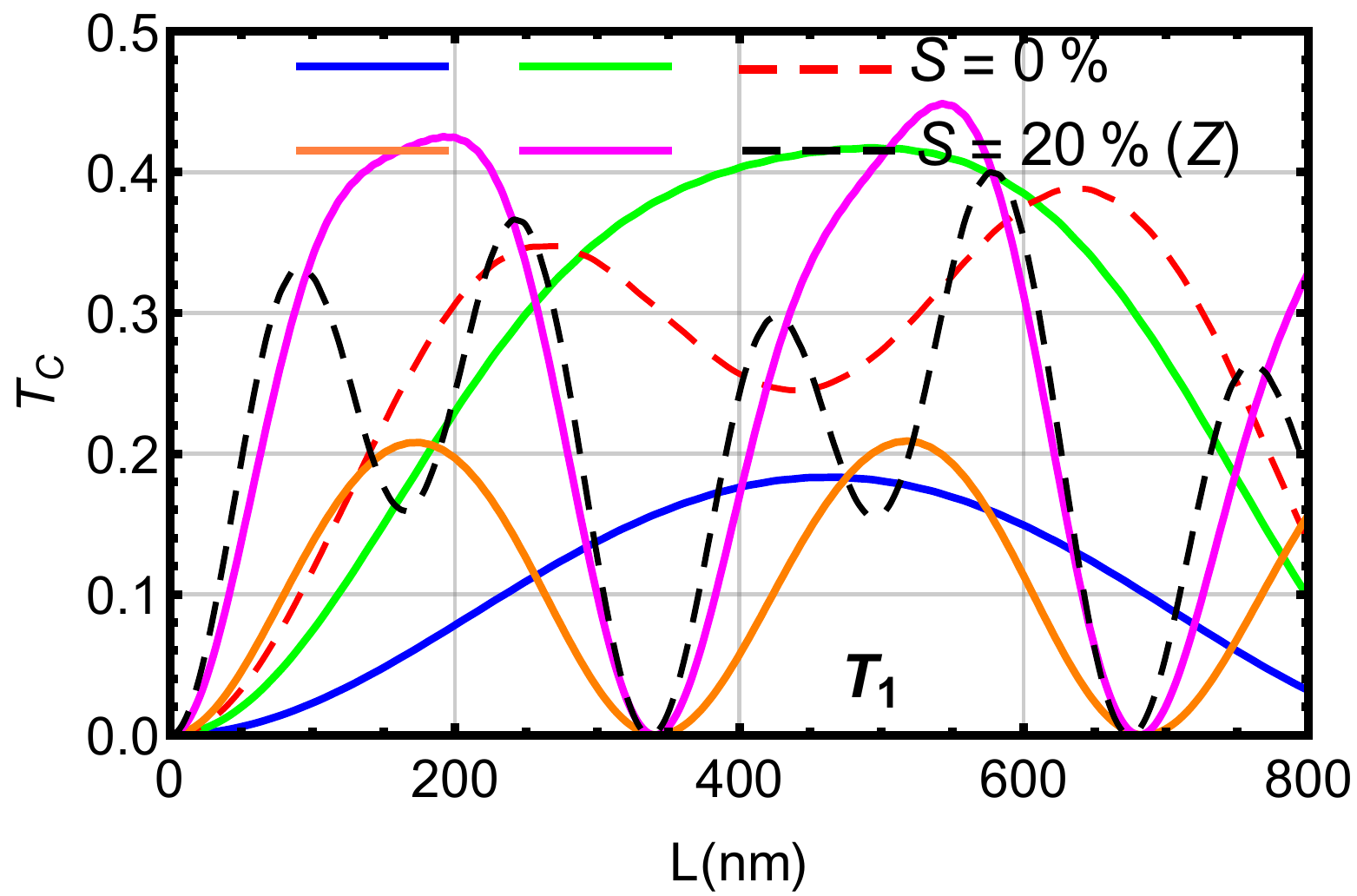}\label{ka1}}
	\subfloat[]{
		\includegraphics[width=0.326\linewidth, height=0.24\textheight]{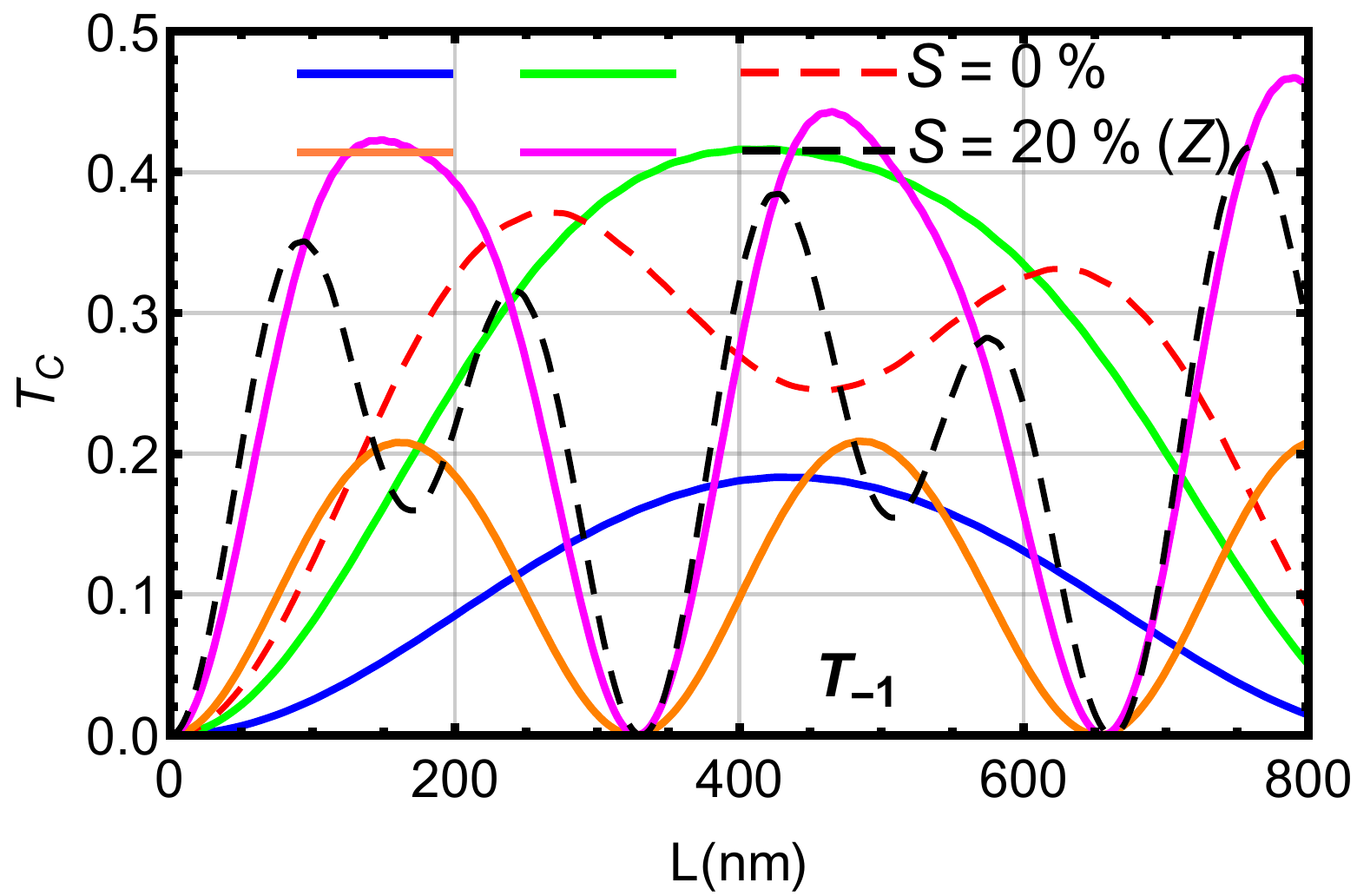}\label{ka2}}
	\caption{{\sf{(color online) Transmission probabilities {of the central band $T_{0}$ and first sidebands $T_{\pm 1}$} 
 as function of the barrier width $L$ at normal incidence $\theta_{0}=0^{\circ}$ for $\mathcal{\varepsilon}=90$ \text{meV}, $\omega=7\times10^{12}$ \text{Hz}, $F=0.015$ \text{V/nm} (blue and orange lines), $F=0.03$ \text{V/nm} (green and magenta lines), $F=0.045$ \text{V/nm} (red and black lines). \textbf{\color{blue}{(a)-(c)}}\color{black}{:} Effect of armchair strain direction for $S=20\%$. \textbf{\color{blue}{(d)-(f)}}\color{black}{:} Effect of zigzag strain direction for $S=20\%$.}}}\label{fiz}
\end{figure}

In order to understand the effect of the frequency $\omega$ and the strain magnitude $S$ on the tunneling transport Dirac fermions through graphene laser barrier at normal incidence $\theta_{0}=0^{\circ}$, we display in {\color{blue}\textbf{Figure}} \ref{fz} the transmission probabilities
as function of the barrier width $L$ for $\mathcal{\varepsilon}=90$ \text{meV},  $\omega=7.5\times10^{12}$ \text{Hz} (blue and orange lines), $\omega=8\times10^{12}$ \text{Hz} (green and magenta lines),  $\omega=8.5\times10^{12}$ \text{Hz} (red and black lines), $F=0.05$ \text{V/nm},  $S=0\%$,  armchair direction $S=6.5\%$ and  zigzag direction  $S=10\%$.  Indeed, we analyze the transmission behavior of the central band for the strainless case ($S=0\%$) by considering three different zones. For the first zone $L$ $\in$ [$0$, $200$ \text{nm}], $T_{0}$ is unity and decreases rapidly for small values of $L$ but it increases by increasing the frequency $\omega$. In the second zone $L$ $\in$ [$200$ \text{nm}, $400$ \text{nm}], we observe that the damped amplitude of $T_{c}$ decreases quickly until it becomes zero as long as $\omega$ increases. For the third zone where $L$ is exceeds the value $400$ \text{nm}, the amplitude of $T_{0}$ increases until it reaches a maximum value. In contrast, one clearly sees  that  the transmission through the first sideband $T_{+1}$ is null at $L=0$ but increases when $L$ increases and coincides in the interval $0<L<{150}$ {\text{nm}} whatever the values taken by $\omega$. After that, we notice the appearance of the oscillations with different amplitudes. The sideband transmission for the photon emission $T_{-1}$ shows the same characteristics of $T_{+1}$ except that it is reversed and their amplitude increases quickly in the interval {$505$} {\text{nm}} $<L<800$ \text{nm}.
When the strain is applied along armchair direction with $S=10\%$ as depicted in {\color{blue}\textbf{Figures}} \ref{fz}\textbf{\color{blue}{(a)}}, \ref{fz}\textbf{\color{blue}{(b)}}, \ref{fz}\textbf{\color{blue}{(c)}}, the transmissions of the central  band $T_{0}$ and  first sidebands $T_{\pm1}$ exhibit a translation to the right and show the same behavior as for $S=0\%$,    in the second zone there is the disappearance of the peaks for the central band transmission.
 Additionally, we notice the emergence of the new oscillations for the first sideband state $T_{+1}$ in the interval $200$ \text{nm} $<L<450$ \text{nm}, which is not the case  
for the transmission via photon emission $T_{-1}$. Now for zigzag direction with strain $S=10\%$ as shown in {\color{blue}\textbf{Figures}} \ref{fz}\textbf{\color{blue}{(d)}}, \ref{fz}\textbf{\color{blue}{(e)}}, \ref{fz}\textbf{\color{blue}{(f)}}, we clearly see the  changes because the transmission over all sidebands oscillates with different amplitude and shifted to the left as well as the number of its oscillations increases in a rapid manner when $S$ and $\omega$ increases. In conclusion, we notice that the strain amplitude $S$ and the frequency $\omega$ play an important role in adjusting transmission behavior. 
\begin{figure}[H]
	\centering
	\subfloat[]{
		\includegraphics[width=0.326\linewidth, height=0.22\textheight]{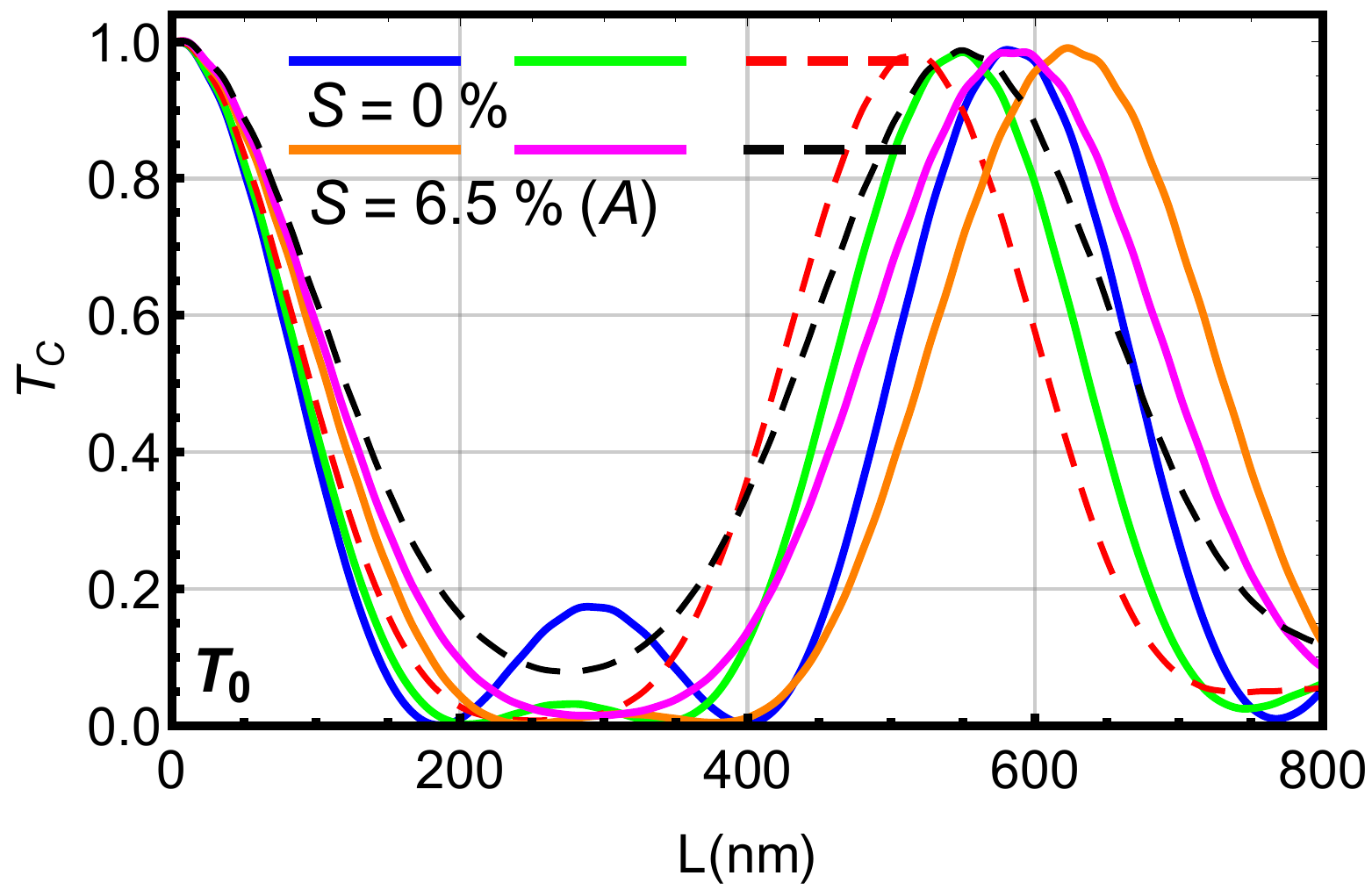}\label{aK}}
	\subfloat[]{
		\includegraphics[width=0.326\linewidth, height=0.22\textheight]{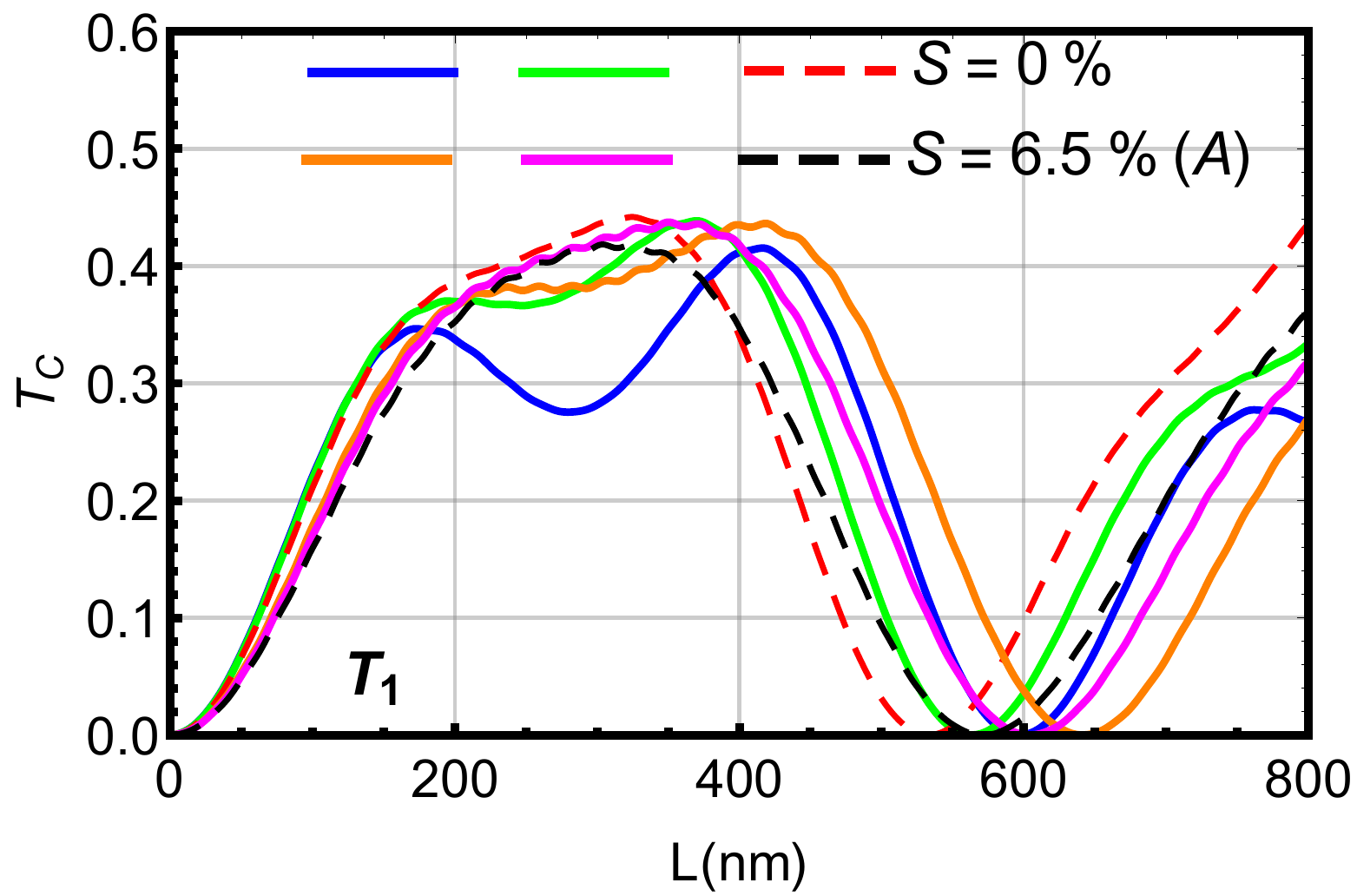}\label{kaL}}
	\subfloat[]{
		\includegraphics[width=0.326\linewidth, height=0.22\textheight]{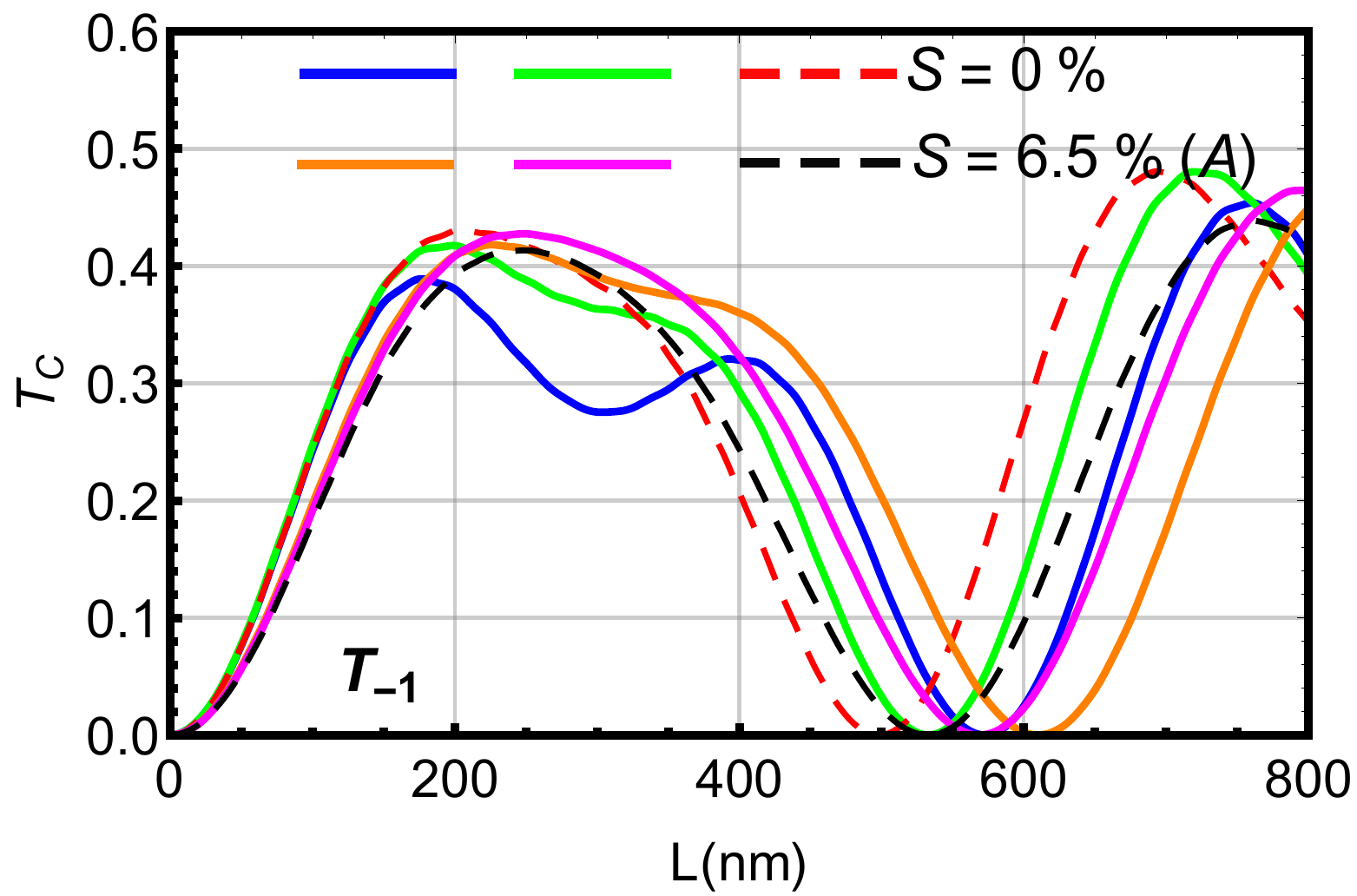}\label{kaM}}
	\\
	\centering
	\subfloat[]{
		\includegraphics[width=0.326\linewidth, height=0.21\textheight]{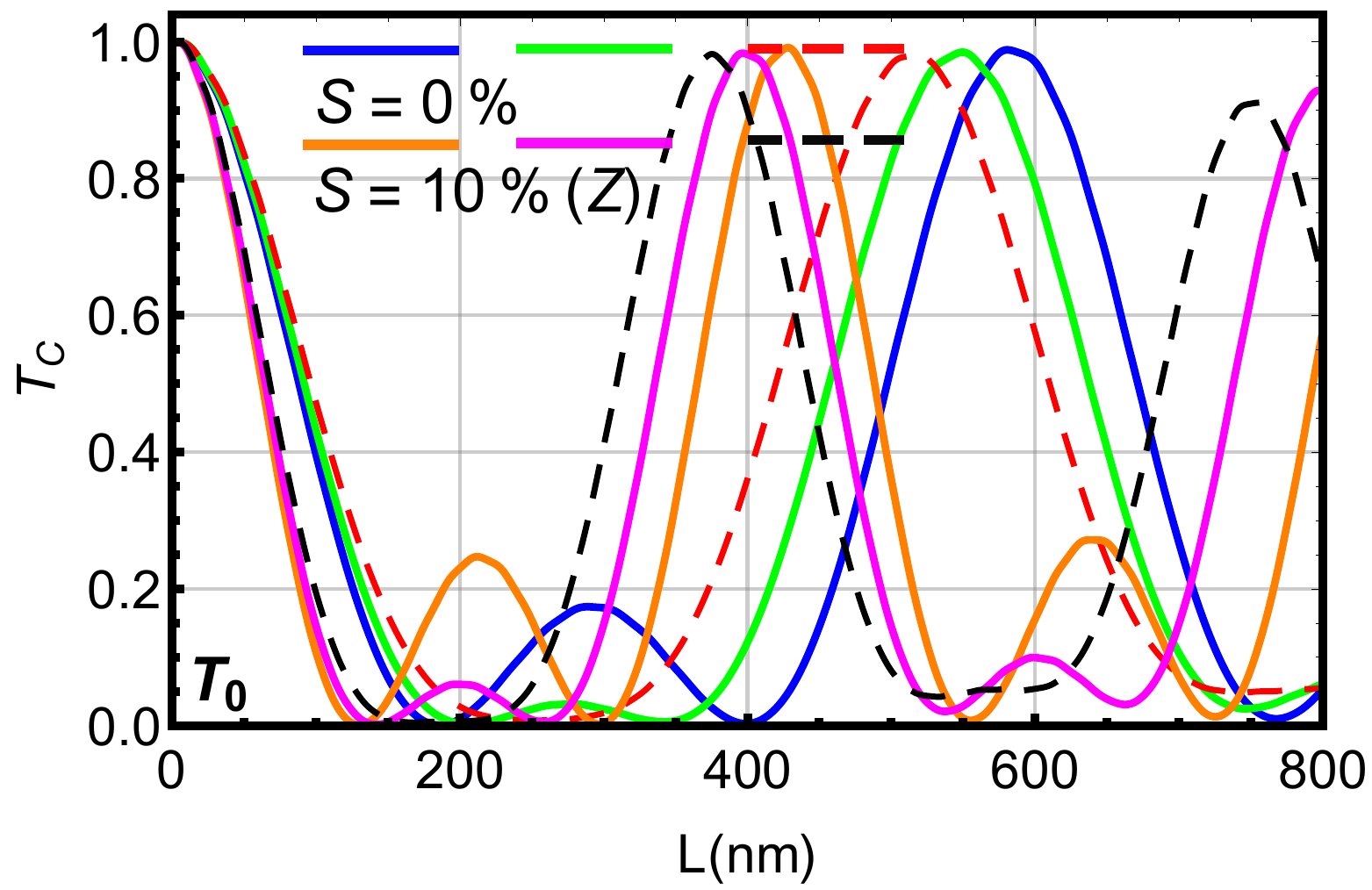}\label{aN}}
	\subfloat[]{
		\includegraphics[width=0.326\linewidth, height=0.21\textheight]{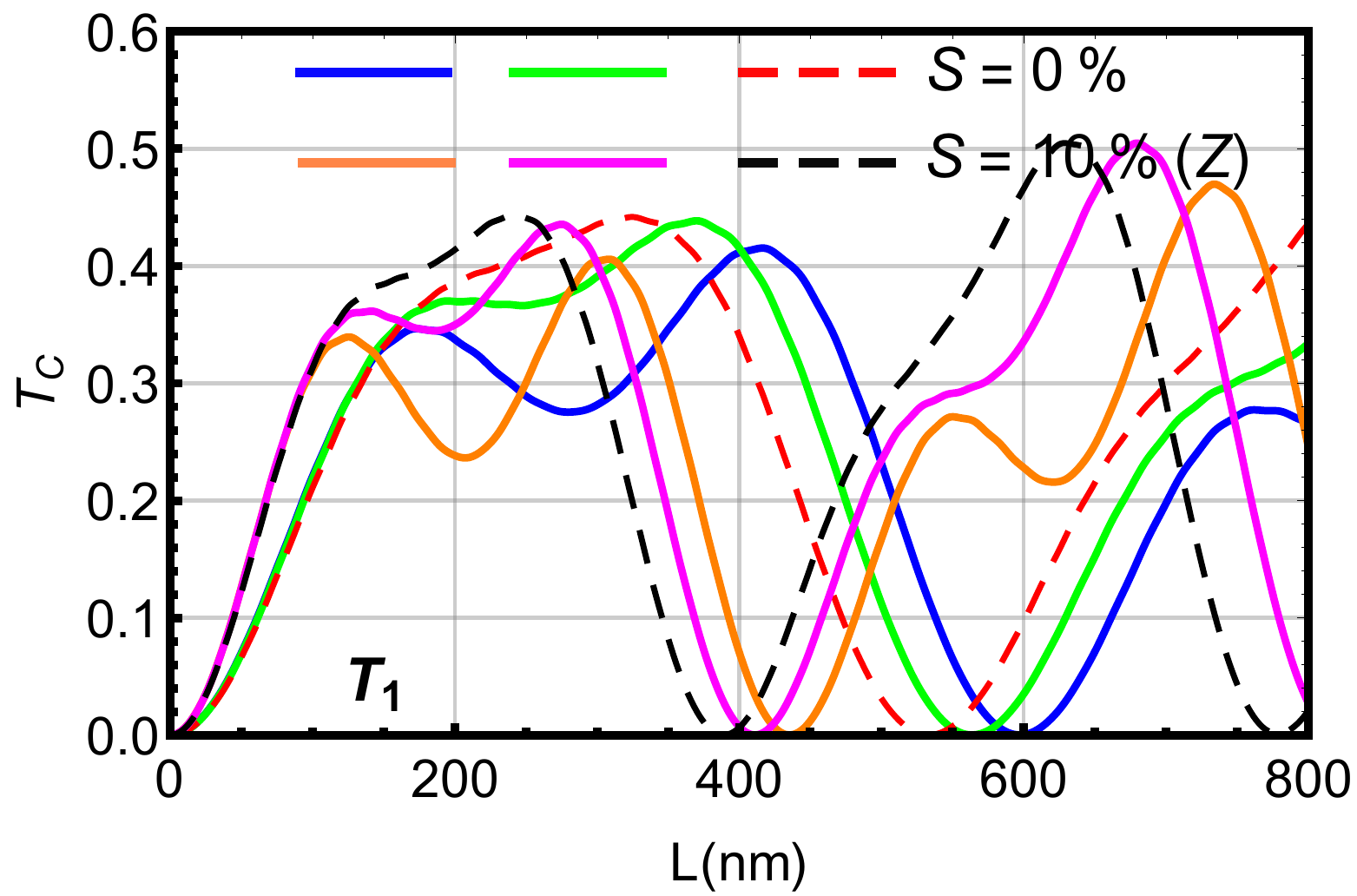}\label{kaN}}
	\subfloat[]{
		\includegraphics[width=0.326\linewidth, height=0.21\textheight]{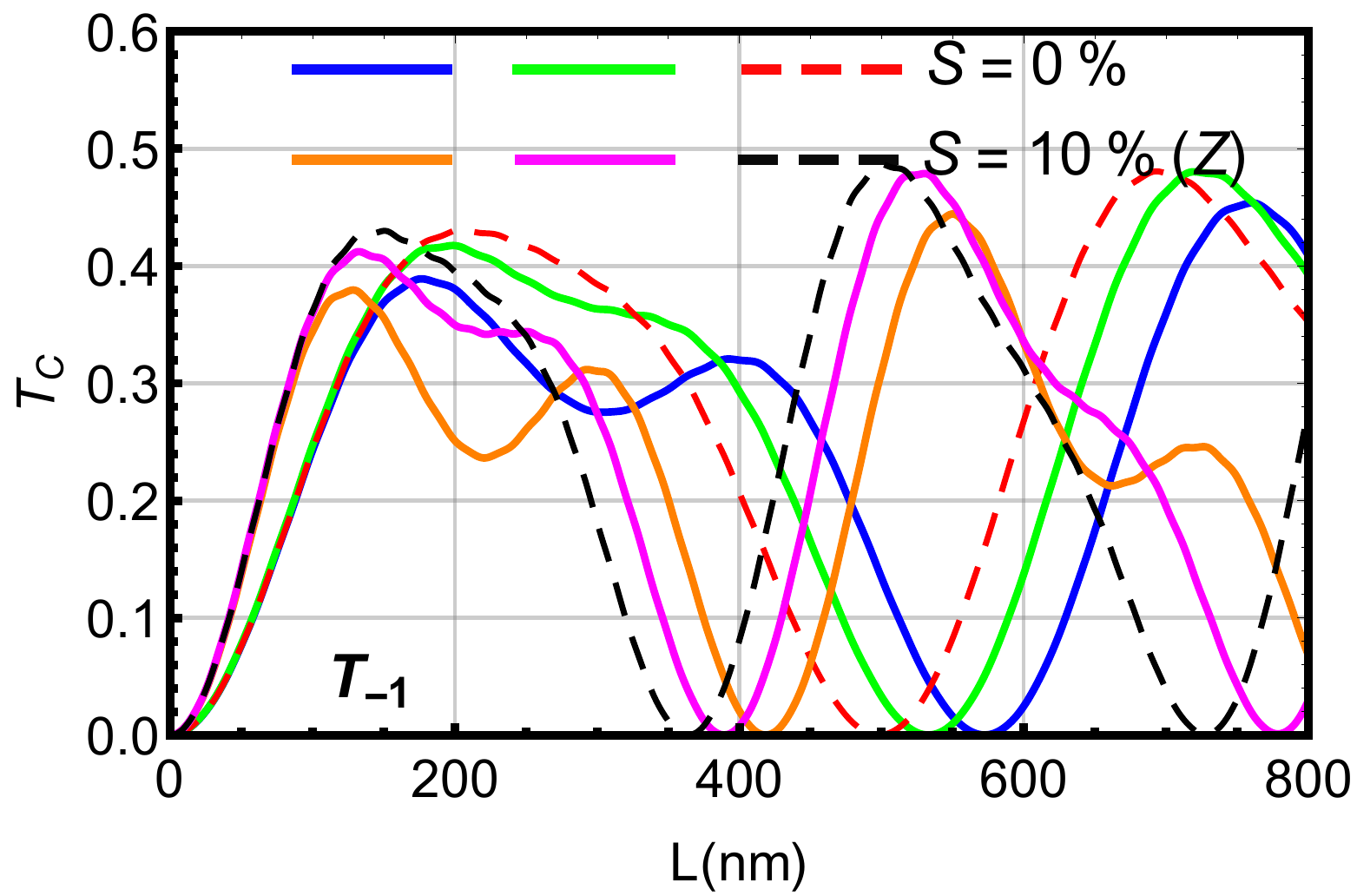}\label{kaI}}
	\caption{{\sf{(color online) Transmission probabilities {of the central band $T_{0}$ and first sidebands $T_{\pm 1}$} 
				as function of the barrier width $L$ at angle incidence $\theta_{0}=20^{\circ}$ for $\mathcal{\varepsilon}=90$ \text{meV}, $\omega=7.5\times10^{12}$ \text{Hz} (blue and orange lines), $\omega=8\times10^{12}$ \text{Hz} (green and magenta lines),  $\omega=8.5\times10^{12}$ \text{Hz} (red and black lines), $F=0.05$ \text{V/nm}.  \textbf{\color{blue}{(a)-(c)}}\color{black}{:} Effect of armchair strain direction for  $S=6.5\%$. \textbf{\color{blue}{(d)-(f)}}\color{black}{:} Effect of zigzag strain direction for $S=10\%$.}}}\label{fz}
\end{figure}

Now, to show the influence of the introduced strain on the total  transmission  probability, we illustrate in {\color{blue}\textbf{Figure}} \ref{fi} $T_{c}$ as function of the barrier width $L$ at normal incidence  $\theta_{0}=0^{\circ}$ for $\mathcal{\varepsilon}=90$ \text{meV}, $\omega=7\times10^{12}$ \text{Hz}, $F=0.05$ \text{V/nm} with $S=0\%$, $S=13\%$, $S=20\%$ are shown in {\color{blue}\textbf{Figures}} \ref{fi}\textbf{\color{blue}{(a)}}, \ref{fi}\textbf{\color{blue}{(b)}}, \ref{fi}\textbf{\color{blue}{(c)}} respectively. Indeed, in the absence of the strain amplitude ($S=0\%$), we observe that the total transmission starts from unit when $L$ is small (close to zero) and takes the form of the sinusoidal function for large values of $L$. When the strain follows armchair direction, we clearly see the disappearance of the oscillations  located in the interval $300$ \text{nm} $<L<600$ \text{nm} for $S=0\%$. Also,  $T_{c}$ {quickly} increases  and displaces to the up as long as $S$ increases {as clearly seen in {\color{blue}\textbf{Figures}} \ref{fi}\textbf{\color{blue}{(b)}}, \ref{fi}\textbf{\color{blue}{(c)}}}. On the other hand, there is significant modification if the deformation is along zigzag direction because we notice the appearance of the Fano type sinusoidal peaks with different amplitudes and it increases by varying $S$. {Then, one may notice that the strain exerted along different directions play the opposite roles on the transmission behaviors} {where the increase of armchair/zigzag deformation makes the oscillations of total channels disappears/appears.}

\begin{figure}[H]
	\centering
	\subfloat[]{
		\includegraphics[width=0.326\linewidth, height=0.24\textheight]{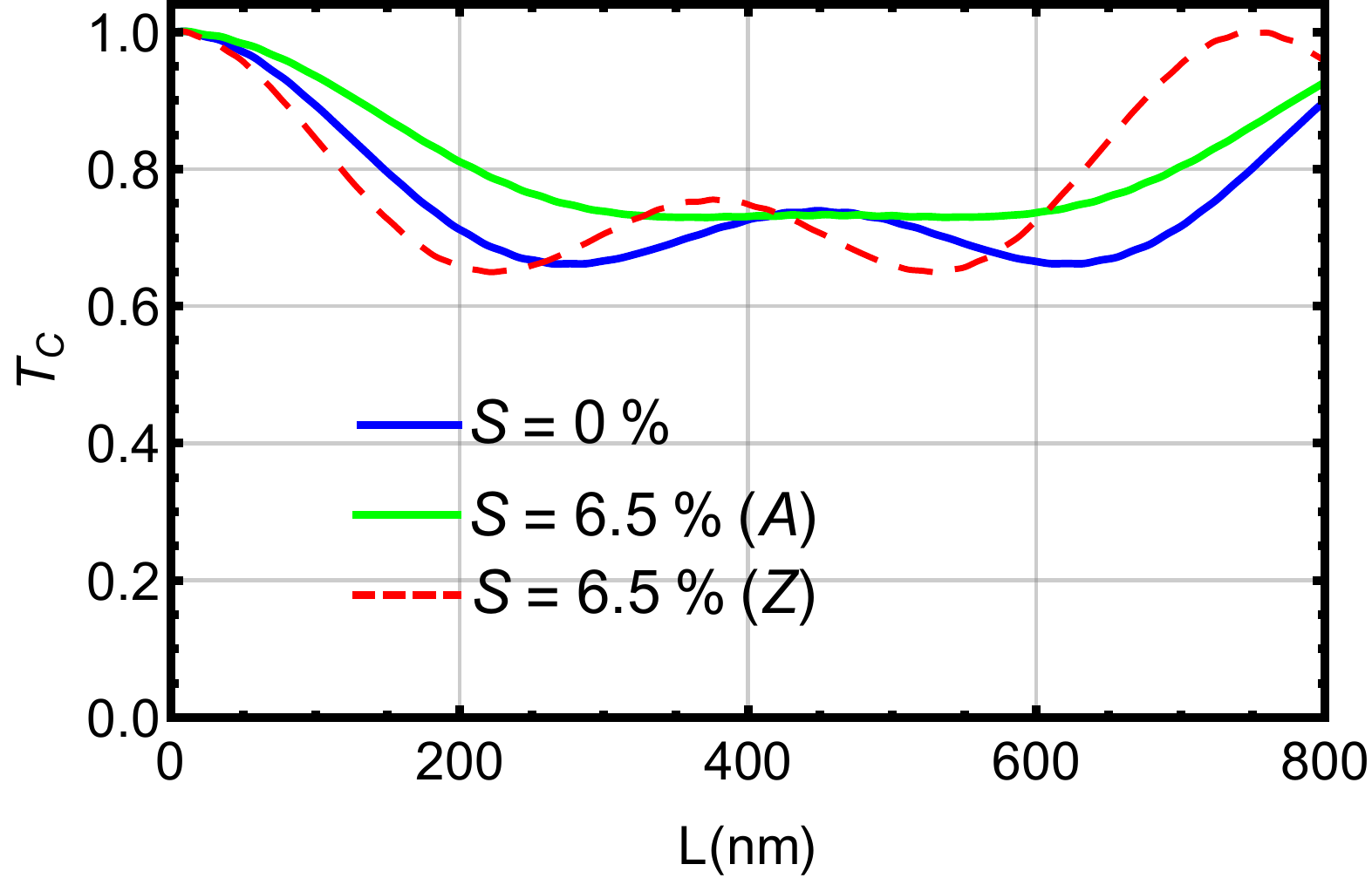}\label{gd}}
	\subfloat[]{
		\includegraphics[width=0.326\linewidth, height=0.24\textheight]{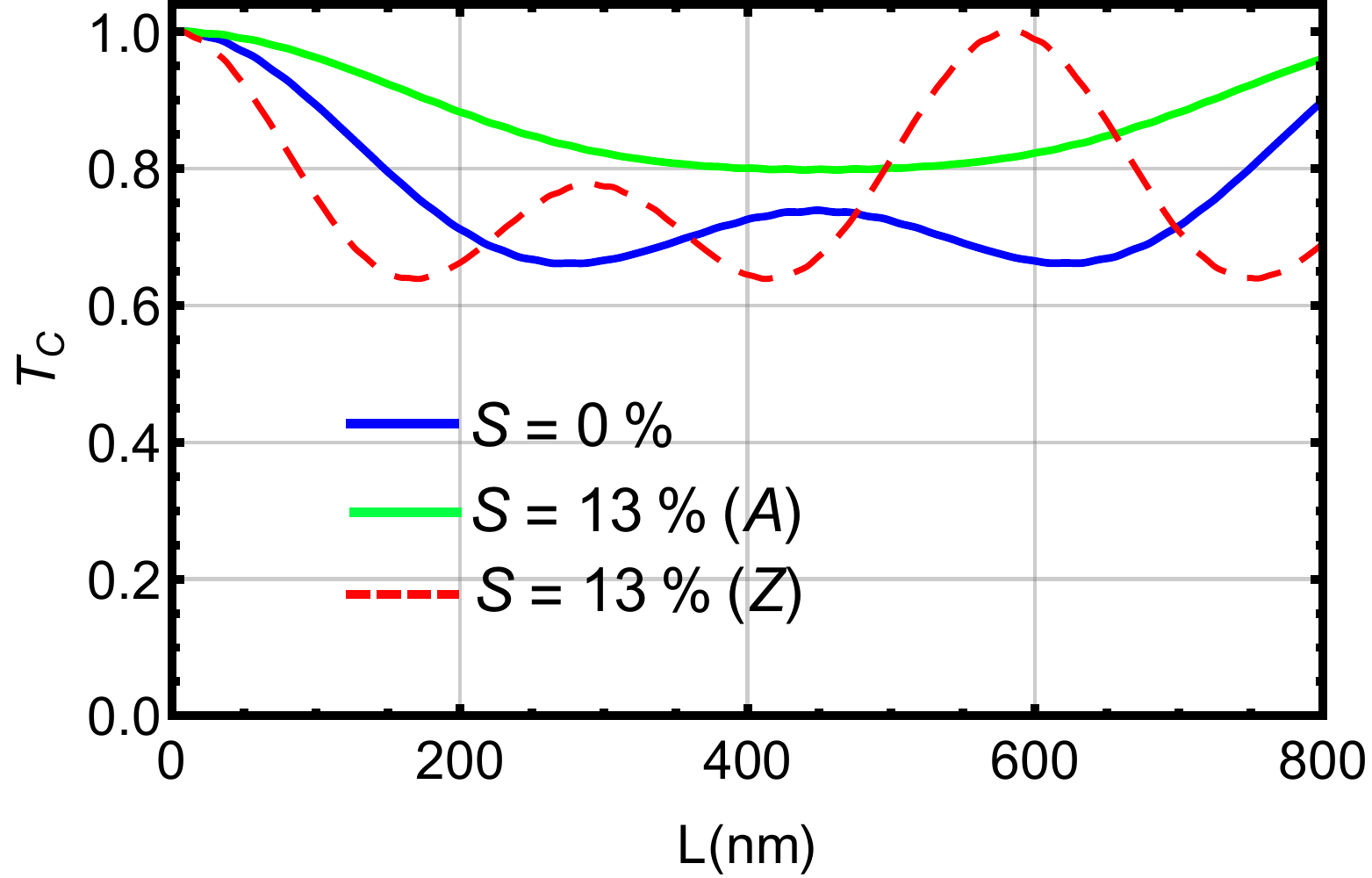}\label{gd1}}
	\subfloat[]{
		\includegraphics[width=0.326\linewidth, height=0.24\textheight]{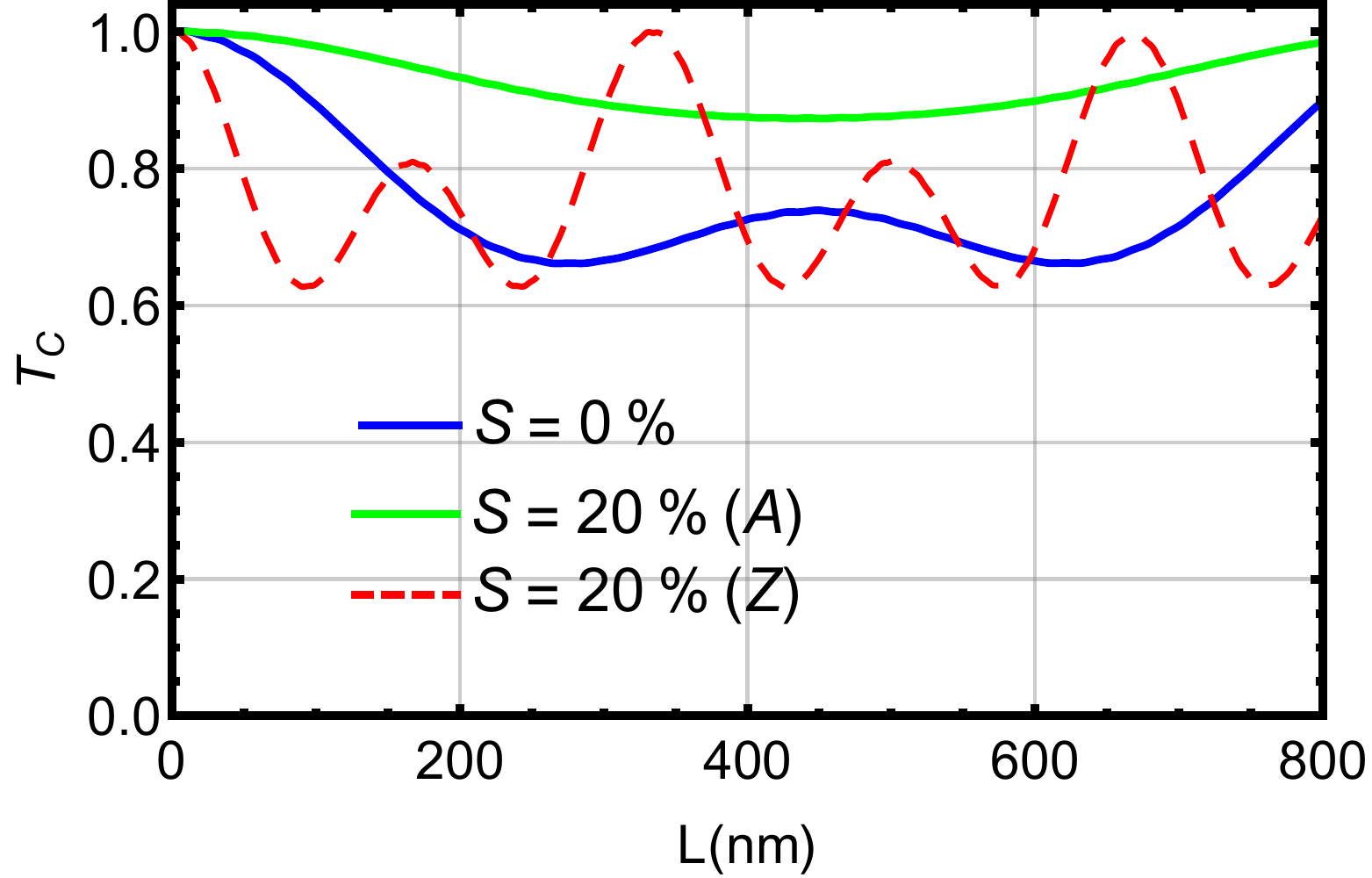}\label{gd2}}	
	\caption{\sf{(color online) Total transmission probability  $T_{c}$  as function of the barrier width $L$ at normal incidence $\theta_{0}=0^{\circ}$ for $\mathcal{\varepsilon}=90$ \text{meV}, $\omega=7\times10^{12}$ \text{Hz}, $F=0.05$ \text{V/nm}, with  strainless $S=0\%$,  armchair direction $S=6.5\%, 13\%, 20\%$, and  zigzag direction $S=6.5\%, 13\%, 20\%$.}}\label{fi}
\end{figure}

To illustrate the combined effects of the laser field amplitude $F$ and strain magnitude $S$ on the tunneling spectra of Dirac fermion in graphene through the laser assisted barrier at normal incidence $\theta_{0}=0^{\circ}$, we plot in {\color{blue}\textbf{Figure}} \ref{fi1}  $T_{c}$ as function of the barrier width $L$ for $\mathcal{\varepsilon}=90$ \text{meV}, $\omega=7\times10^{12}$ \text{Hz} and three values of the laser field amplitude $F=0.035$ \text{V/nm} (blue line), $F=0.045$ \text{V/nm} (green line), $F=0.055$ \text{V/nm} (red line) with $S=0\%$, $S=20\%$ (A), $S=20\%$ (Z). According to {\color{blue}\textbf{Figure}} \ref{fi1}\textbf{\color{blue}{(a)}}, we observe that for the strainless case ($S=0\%$) $T_{c}$ changes slowly closed to the unit and significantly decreases by increasing  $F$. Importantly, one sees that for $F=0.055$ \text{V/nm} (red line) $T_{c}$ can be accompanied by sinusoidal function with a clear amplitude localized in the zone $L$ $\in$ [$250$ \text{nm}, $650$ \text{nm}]. Regarding the strain along armchair direction with $S=20\%$,  {\color{blue}\textbf{Figure}} \ref{fi1}\textbf{\color{blue}{(b)}} shows the same behavior as in {\color{blue}\textbf{Figure}} \ref{fi1}\textbf{\color{blue}{(a)}} except that $T_{c}$ displaces to the up and their oscillations disappear. Now for zigzag direction with $S=20\%$ in {\color{blue}\textbf{Figure}} \ref{fi1}\textbf{\color{blue}{(c)}}, we observe that the number of oscillations and their amplitude increase dramatically as long as $F$ increases. It is clearly seen that there is appearance of the periodic sinusoidal oscillations at $F=0.035/0.055$ \text{V/nm}. We conclude that the obvious amendment on total transmission behavior depends on different the strain  and   laser field.
\begin{figure}[H]
	\centering
	\subfloat[]{
		\includegraphics[width=0.326\linewidth, height=0.24\textheight]{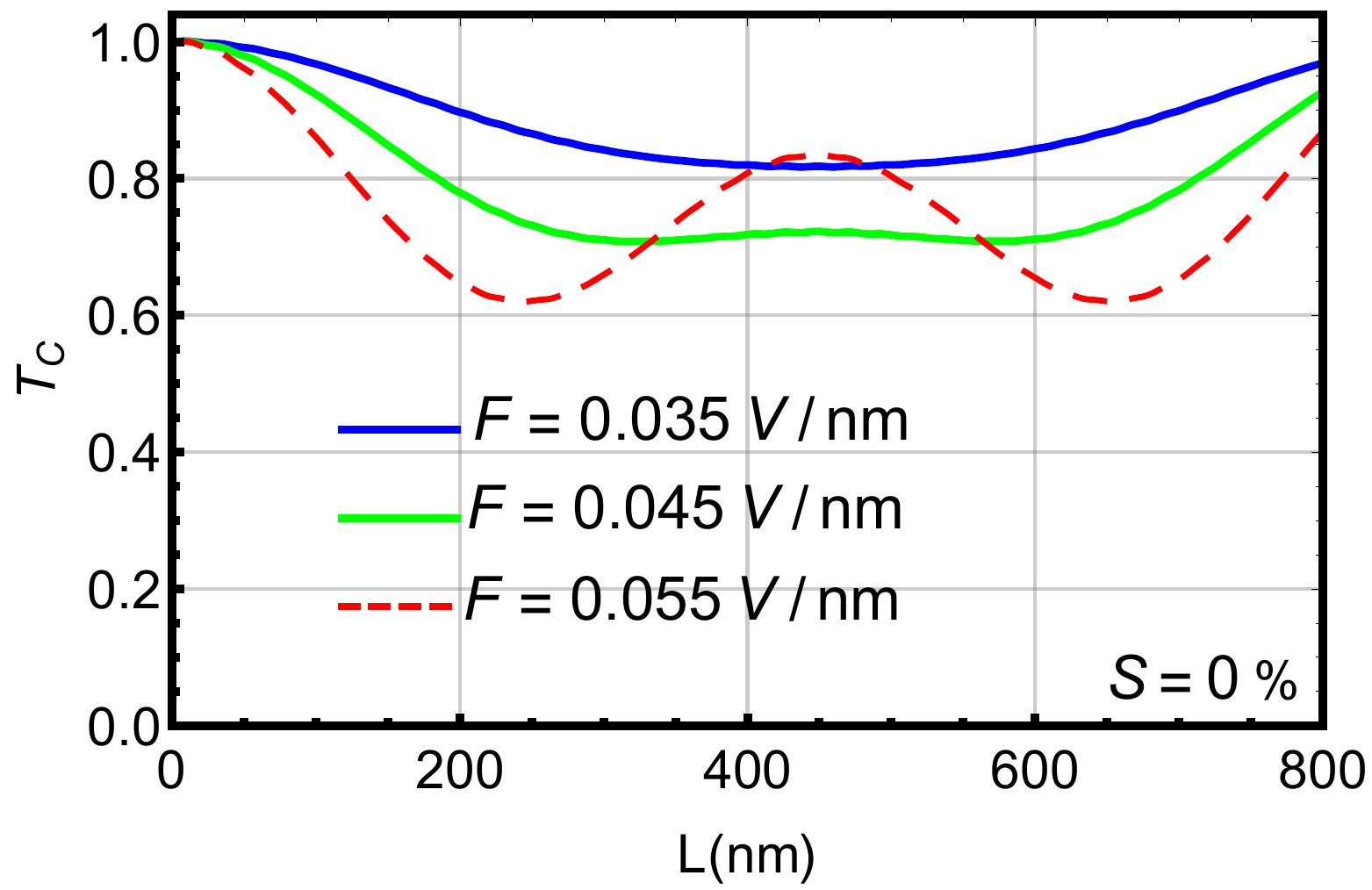}\label{Kd}}
	\subfloat[]{
		\includegraphics[width=0.326\linewidth, height=0.24\textheight]{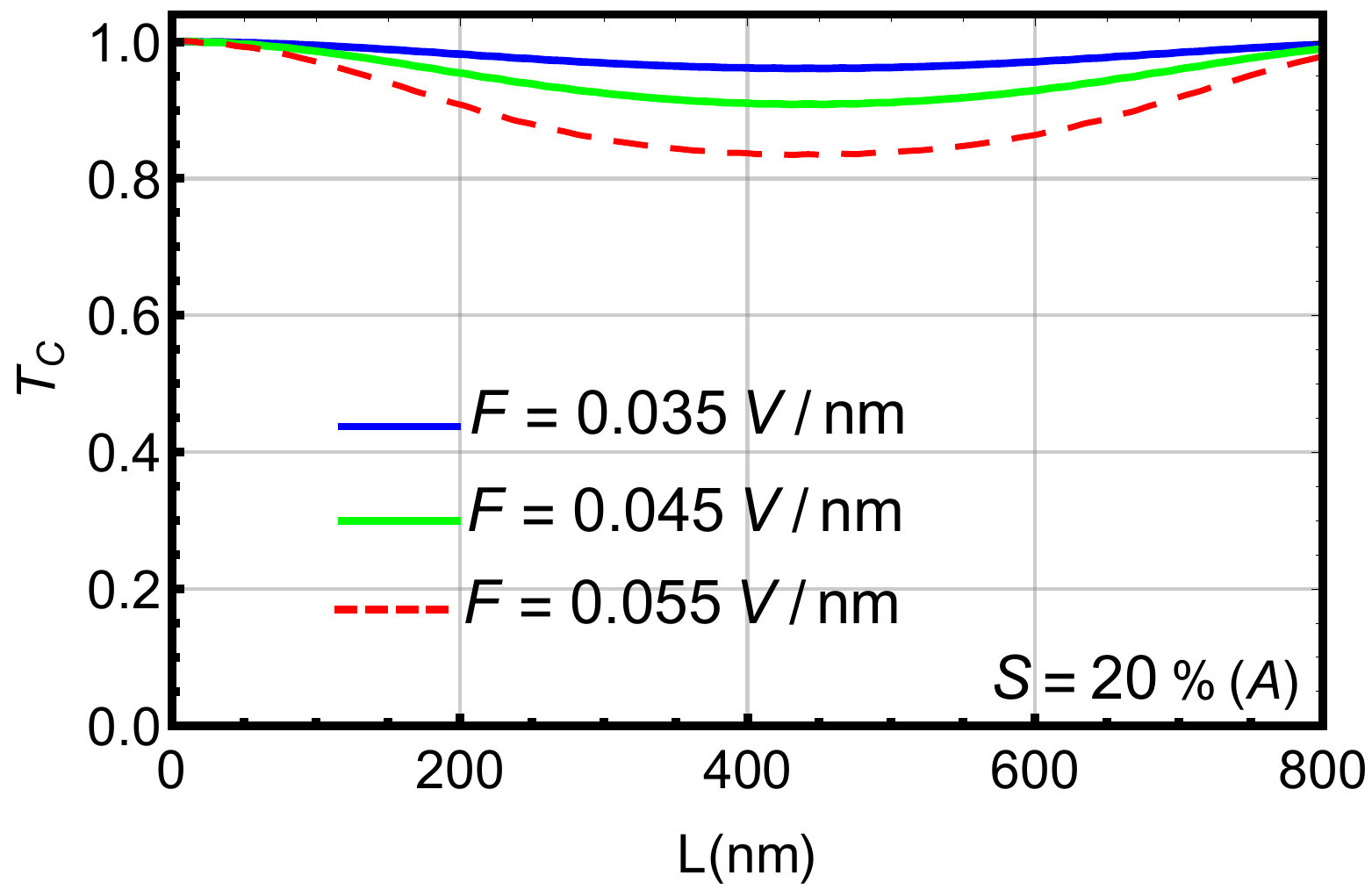}\label{Kd1}}
	\subfloat[]{
		\includegraphics[width=0.326\linewidth, height=0.24\textheight]{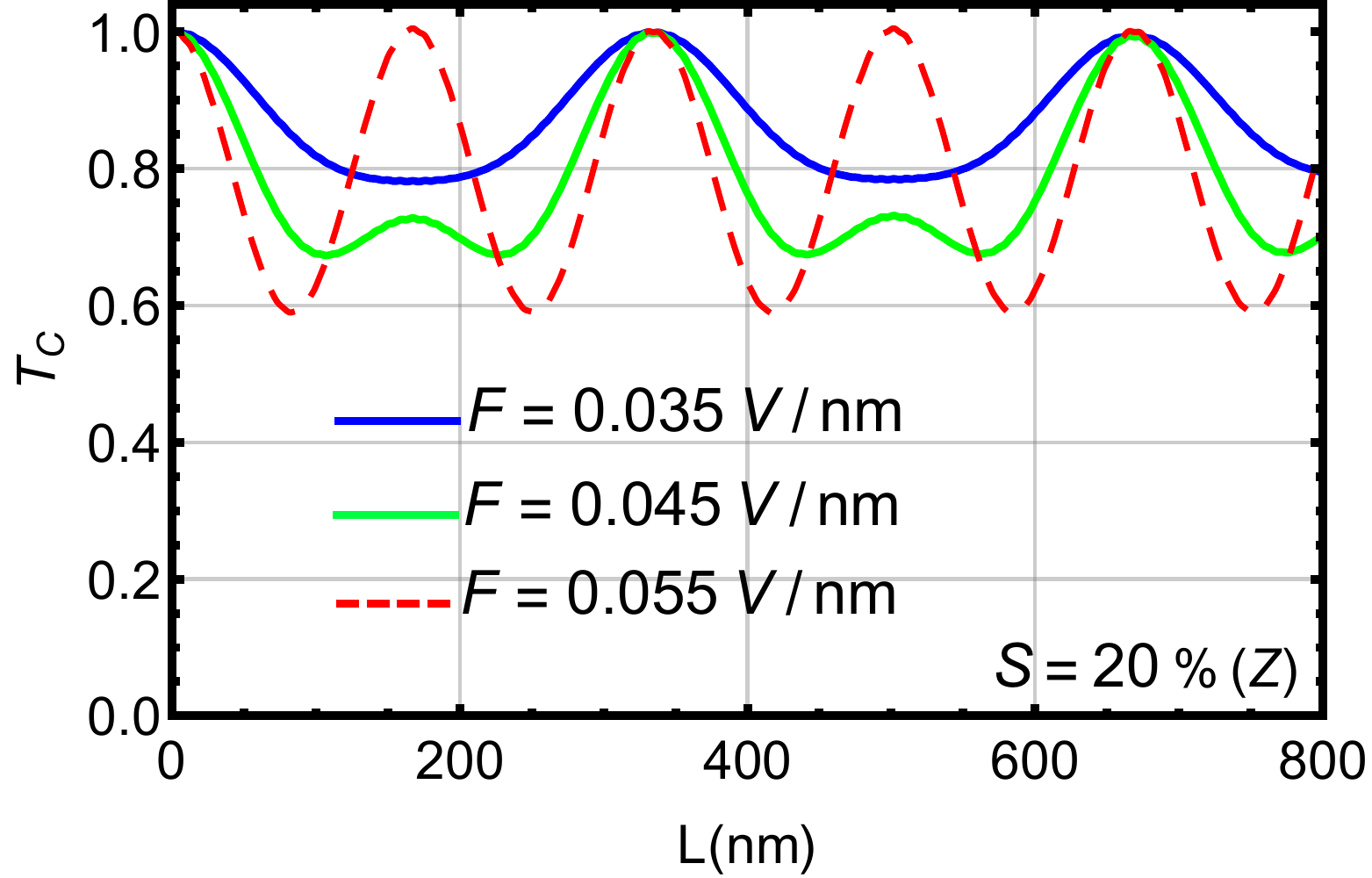}\label{Kd2}}	
\caption{\sf{(color online) Total transmission probability $T_{c}$  as function of the barrier width $L$ at normal incidence $\theta_{0}=0^{\circ}$ for $\mathcal{\varepsilon}=90$ \text{meV}, $\omega=7\times10^{12}$ \text{Hz}, {$F=0.035$ \text{V/nm} (blue line), $F=0.045$ \text{V/nm} (green line), $F=0.055$ \text{V/nm} (red line)}, \textbf{\color{blue}{(a)}}\color{black}{:} Without strain $S=0\%$. \textbf{\color{blue}{(b)}}\color{black}{:} Effect of armchair strain direction for $S=20\%$. \textbf{\color{blue}{(c)}}\color{black}{:} Effect of zigzag strain direction for $S=20\%$.}}\label{fi1}
\end{figure} 
\section{Conclusion}

{We have theoretically studied the transport properties 
through normal/strained/normal graphene subjected to a linearly polarized dressing field. After solving down Dirac equation, we have  determined the eigenvalues and corresponding eigenspinors in each region composing our system.} 
{Using} the boundary condition with transfer matrix method and Floquet approach to derive the transmission {probability} in the different Floquet sideband states as function of the physical parameters characterizing our system. 

Furthermore, we have numerically analyzed 
{the transmission probability for two directions of strain including zigzag and armchair} under suitable conditions of the barrier width, laser field amplitude, frequency and strain amplitude. 
Indeed, we have observed that in the case of without strain {($S=0\%$)} the transmission of the central band {reduces} for slight values of the barrier width { but has a damped oscillatory profile for its first strong values and after that it grows slowly}. {While} the transmission  {of} {the other sidebands starts increasing} from zero and oscillates  {with a different amplitude.  

As a results,} it was shown that at normal incidence {($\theta_{0}=0^{\circ}$) the transmission of} photon absorption 
{does not coincide} than the process {of} photon emission 
{as long as the barrier width increases unlike} the case of oscillating barrier {in time}.
{ On other hand, it is noticed that} 
{for positive angle incidence ($\theta_{0}>0^{\circ}$) the oscillation frequency of transmission is higher than for negative angle incidence ($\theta_{0}<0^{\circ}$)}. 
{However, when the strain magnitude is switched on ($S\neq0\%$) 
the number of peaks in all transmission channels decreases as long as the strength of armchair direction but increases strongly for 
zigzag case. In addition}, it is found that the { amplitude and the frequency of the laser field} affect $T_{c}$ where we have observed the appearance of Fano type resonance oscillations. 

 \section*{Acknowledgment}
The generous  support provided by the Saudi Center for Theoretical Physics (SCTP) is highly appreciated by all authors.



\begin{thebibliography}{99}
\bibitem{q1} K. S. Novoselov, A. K. Geim, S. V. Morozov, D. Jiang, Y. Zhang, S. V. Dubonos, I. V. Grigorieva, and A. A. Firsov, Science \textbf{306}, 666 (2004).
\bibitem{q2} K. S. Novoselov, A. K. Geim, S. V. Morozov, D. Jiang, Y. Zhang, S. V. Dubonos, I. V. Grigorieva, and A. A. Firsov, Nature \textbf{438}, 197 (2005).
\bibitem{q5} A. H. C. Neto, F. Guinea, N. M. R. Peres, K. S. Novoselov, and A. K. Geim, Rev. Mod. Phys. \textbf{81}, 109 (2009).
\bibitem{q6} E. McCann and V. I. Fal'ko, Phys. Rev. Lett. \textbf{96}, 086805 (2006). 
\bibitem{q7} Y. B. Zhang, Y. W. Tan, H. L. St\"ormer, and P. Kim, Nature \textbf{438}, 201 (2005).
\bibitem{q8} V. P. Gusynin and S. G. Sharapov, Phys. Rev. Lett. \textbf{95},  146801 (2005).
\bibitem{q9} V. M. Pereira, A. C. Neto, and N. M. R. Peres, Phys. Rev. B \textbf{80}, 045401 (2009).
\bibitem{q10} M. A. H. Vozmedianoa, M. I. Katsnelson, and F. Guinea, Phys.
Rep. \textbf{496}, 109 (2010).
\bibitem{q11} S. M. Choi, S. H. Jhi, and Y. W. Son, Phys. Rev. B \textbf{81}, 081407 (2010).
\bibitem{q12} H. Goudarzi, M. Khezerlou, and H. Kamalipour, Superlattice Microstructure \textbf{83}, 101 (2015).
\bibitem{q13} R. V. Gorbachev, A. S. Mayorov, A. K. Savchenko, D. W. Horsell, and F. Guinea, Nano Lett. \textbf{8}, 1995 (2008). 
\bibitem{q15} C. W. J. Beenakker, Rev. Mod. Phys. \textbf{80}, 1337 (2008). 
\bibitem{qa15} N. Stander, B. Huard, and D. Goldhaber-Gordon, Phys. Rev. Lett. \textbf{102}, 026807 (2009).
\bibitem{q16} X. Chen and J. W. Tao, Appl. Phys. Lett. \textbf{94}, 262102 (2009).
\bibitem{q17} J. J. M. Pereira, P. Vasilopoulos, and F. M. Peeters, Appl. Phys. Lett. \textbf{90}, 132122 (2007). 
{\bibitem{S2} Z. H. Ni, T. Yu, Y. H. Lu, Y. Y. Wang, Y. P. Feng, and Z. X. Shen, ACS Nano. \textbf{2}, 2301 (2008).
\bibitem{S3} J. Lu, A. H. C. Neto, and K. P. Loh, Nat. Commun. \textbf{3}, 823
(2012).
\bibitem{S4} H. Suzuura and T. Ando, Phys. Rev. B \textbf{65}, 235412
(2002).
\bibitem{S5} F. Guinea, M. I. Katsnelson, and A. K. Geim, Nat. Phys. \textbf{6}, 30 (2010).}
{\color{black}\bibitem{ref25} B. Soodchomshom, Physica B \textbf{406}, 614 (2011).
\bibitem{a5} B. Soodchomshom and P. Chantngarm, J. Supercond. Nov. Magn. \textbf{24}, 1885 (2011).
\bibitem{a7} J. H. Wong, B. R. Wu, and M. F. Lin, J. Phys. Chem. C \textbf{116}, 8271  (2012).
\bibitem{Xu} K. Dini, O. V. Kibis, and I. A. Shelykh, Phys. Rev. B \textbf{93}, 235411 (2016). 
\bibitem{Xv} F. K. Joibari, Y. M. Blanter, and G. E. W. Bauer, Phys. Rev. B \textbf{90}, 155301 (2014).}
\bibitem{gt} W. Li and  L. E. Reichi, Phys. Rev. B \textbf{60}, 15732 (1999).
{\bibitem{ly1} F. K. Joibari, Y. M. Blanter, and G. E. W. Bauer, Phys. Rev. B \textbf{90}, 155301 (2014).
\bibitem{ly2} K. L. Koshelev, V. Y. Kachorovskii, and M. Titov, Phys. Rev. B \textbf{92}, 235426 (2015).}
\bibitem{Xu1} T. Oka and H. Aoki, Phys. Rev. B \textbf{79}, 081406 (2009).
\bibitem{Xu2} O. V. Kibis, O. Kyriienko, and I. A. Shelykh, Phys. Rev. B \textbf{84}, 195413 (2011).
\bibitem{strain} H. Chnafa, M. Mekkaoui, A. Jellal, and A. Bahaoui, Eur. Phys. J. B \textbf{94}, 39 (2021).
\bibitem{Xu3} J. T. Liu, F. H. Su, H. Wang, and X. H. Deng, Europhys. Lett. \textbf{95}, 24003 (2011).
\bibitem{XU1} V. Ryzhii and M. Ryzhii, Phys. Rev. B \textbf{79}, 245311 (2009).
\bibitem{XU2} A. R. Wright, J. C. Cao, and C. Zhang, Phys. Rev. Lett. \textbf{103}, 207401 (2009).
\bibitem{f1} C. Sinha and R. Biswas, Appl. Phys. Lett. \textbf{100}, 183107 (2012).
\bibitem{Xu4} R. Biswas and C. Sinha, Appl. Phys. \textbf{114}, 183706 (2013).
{\color{black}\bibitem{Xu5} R. Biswas, S. Maity, and C. Sinha, Physica E \textbf{84}, 235 (2016).}  
\bibitem{a6} W. Yan, Physica B \textbf{504}, 23 (2017).
\bibitem{a8} W. X. Yan and L. N. Ma, Physica B \textbf{28}, 445 (2014).
\bibitem{yu} H. M. Sadd, Elasticity: theory, applications, and numerics. (Academic Press, 2009).
\bibitem{Dipole} R. Loudon, The Quantum Theory of Light, 3rd ed. (Oxford University Press Inc., New York, 2000).
\bibitem{a1} Y. B. Zel’Dovich, Sov. Phys. JETP \textbf{24}, 1006 (1967).
\bibitem{a3} M. Grifoni and P. H\"{a}nggi, Phys. Rep. \textbf{304}, 229 (1998).
{\bibitem{a4} G. Platero and R. Aguado, Phys. Rep. \textbf{395}, 1 (2004).}
{\bibitem{ai} Z. Gu, H. A. Fertig, D. P. Arovas, and A. Auerbach, Phys. Rev. Lett. \textbf{107}, 216601 (2011).}
\bibitem{X8} M. A. Zeb, K. Sabeeh, and M. Tahir, Phys. Rev. B \textbf{78}, 165420 (2008).
{\bibitem{ln} B. Trauzettel, Y. M. Blanter, and A. F. Morpurgo, Phys. Rev. B \textbf{75}, 035305 (2007).}
\bibitem{lp} E. B. Choubabi, A. Jellal, and M. Mekkaoui, Eur. Phys. J. B \textbf{92}, 85 (2019).
\bibitem{lo} B. Lemaalem, M. Mekkaoui, A. Jellal, and H. Bahlouli, Eur. Phys. Lett. \textbf{129}, 27001 (2020).
\bibitem{fb1} M. Barbier, P. Vasilopoulos, and F. M. Peeters, Phys.
Rev. B \textbf{82}, 235408 (2010).
\bibitem{fa1} M. Barbier, P. Vasilopoulos, F. M. Peeters, and J. M. Pereira, Phys. Rev. B \textbf{79}, 155402 (2009).
{\color{black}\bibitem{f2} A. Jellal, M. Mekkaoui, E. B. Choubabi, and H. Bahlouli, Eur. Phys. J. B \textbf{87}, 123 (2014).
\bibitem{gt1} C. Sinha and R. Biswas, Phys. Rev. B \textbf{84}, 155439 (2011).}
\end{thebibliography}
\end{document}